\pdfoutput=1
\documentclass[11pt]{article}
\usepackage[journal=jquant,stage=submission,logging=stdout]{jsty3-author}

\usepackage[utf8]{inputenc}
\usepackage{microtype}
\usepackage{amsthm,mathtools,bm}
\usepackage[nameinlink]{cleveref}
\crefname{equation}{eq.}{eqs.}
\Crefname{equation}{Eq.}{Eqs.}
\crefname{figure}{figure}{figures}
\crefname{section}{section}{sections}
\crefname{appendix}{appendix}{appendices}

\hypersetup{pdftitle={Pauli spectrum and nonstabilizerness of random fermionic Gaussian states}}

\newcommand{\E}{\mathbb{E}}
\newcommand{\Pcal}{\mathcal{P}}
\newcommand{\dd}{\,\mathrm d}
\newcommand{\Tr}{\operatorname{Tr}}
\newcommand{\Pf}{\operatorname{Pf}}
\newcommand{\Law}{\operatorname{Law}}
\newcommand{\diag}{\operatorname{diag}}
\newcommand{\Var}{\operatorname{Var}}
\newcommand{\id}{\mathbb I}
\newcommand{\Gam}{\Gamma}
\newcommand{\loge}{\log_2e}
\newcommand{\rising}[2]{(#1)_{#2}}

\title{Pauli spectrum and nonstabilizerness of random fermionic Gaussian states}

\author*{a}{Xhek Turkeshi}{turkeshi@thp.uni-koeln.de}{0000-0003-1093-3771}
\author{b}{Piotr Sierant}{}{0000-0001-9219-7274}
\author{c,d}{Poetri Sonya Tarabunga}{}{0000-0001-8079-9040}

\affiliation{a}{Institut f\"ur Theoretische Physik, Universit\"at zu K\"oln, Z\"ulpicher Stra{\ss}e 77, 50937 K\"oln, Germany}
\affiliation{b}{Barcelona Supercomputing Center, Pla\c{c}a Eusebi G\"uell 1-3, 08034 Barcelona, Spain}
\affiliation{c}{Technical University of Munich, TUM School of Natural Sciences, Physics Department, James-Franck-Stra{\ss}e 1, 85748 Garching, Germany}
\affiliation{d}{Munich Center for Quantum Science and Technology (MCQST), Schellingstra{\ss}e 4, 80799 M\"unchen, Germany}

\abstract{
We characterize the magic of random fermionic Gaussian states through their Pauli spectrum and their stabilizer entropies.
For the Haar ensemble of Majorana Gaussian states we derive closed finite-$N$ expressions for the average stabilizer purities and reconstruct the Pauli spectrum exactly, as a mixture of products of beta-distributed variables resolved by Majorana weight.
The stabilizer entropies, and their filtered versions, 
freeze at R\'enyi index $q_c=2$: for $q>2$ the magic density is $1/(q-1)$, controlled by rare weight-two Majorana strings. Resolving these contributions requires exponentially many characteristic samples, making direct sampling estimates of the stabilizer entropies inefficient in this frozen regime. For Haar-random Gaussian states at fixed filling, we express the averaged stabilizer purities at positive integer $q$ as coefficient integrals over $\mathrm{SU}(2q)$, with dimension independent of $N$, and evaluate the $q=2$ case exactly.
We establish that the frozen density describes typical states in the Majorana and half-filled number-conserving ensembles and relate these ensembles to the corresponding $\mathrm{SYK}_2$ ground states.  Finally, we show that typical fermionic Gaussian states retain certified magic after discarding any fixed fraction $\kappa<\kappa_\star^{\rm G}\simeq0.7613$ of their modes in the thermodynamic limit, remaining magical even beyond the $2/3$ threshold for Haar-random states in the full Hilbert space.
}
\AtBeginDocument{\nolinenumbers}

\begin{document}

\section{Introduction}
\label{sec:introduction}

Understanding magic in quantum many-body systems is the subject of intense current interest, motivated by both fundamental questions and applications to quantum technologies~\cite{ChitambarGour2019,LiuWinter2022,NiroulaEtAl2024,TurkeshiTirritoSierant2025}. Magic, or nonstabilizerness, is the resource that, when supplied to Clifford circuits, enables universal quantum computation~\cite{BravyiKitaev2005,VeitchMousavianGottesmanEmerson2014,HowardCampbell2017}; it governs both the classical cost of simulating quantum states~\cite{Gottesman1998,AaronsonGottesman2004,BravyiGosset2016} and the fault-tolerant overhead required to prepare them~\cite{BravyiKitaev2005,HowardCampbell2017}. In many-body systems, magic is conveniently quantified by stabilizer R\'enyi entropies (SREs)~\cite{LeoneOliveiroHamma2022}. For R\'enyi index $q\geq2$, SREs are magic monotones~\cite{LeoneBittel2024,HaugPiroli2023}; they can be computed in state-vector simulations~\cite{Sierant26comp,Huang26fast,Xiao26expo} or using Pauli sampling and tensor-network methods~\cite{HaugPiroliMPS2023,LamiCollura2023,TarabungaTirritoChandaDalmonte2023,tarabunga2024mps,tarabunga2025efficientmutualmagicmagic} and measured on quantum hardware~\cite{OlivieroLeoneHammaLloyd2022,HaugLeeKim2024,NiroulaEtAl2024}. 
A finer characterization of magic is provided by the Pauli spectrum, namely the full distribution of Pauli expectation values~\cite{BeverlandCampbellHowardKliuchnikov2020,TurkeshiDymarskySierant2025}.

Matchgate circuits and free-fermion systems~\cite{Valiant2002,TerhalDiVincenzo2002,Knill2001,JozsaMiyake2008} form another class of classically tractable quantum models.
Although matchgate circuits can generate arbitrary amounts of entanglement, they remain efficiently simulable~\cite{Bravyi2005,SuraceTagliacozzo2022}, and hence useful for validation benchmarks of Clifford and doped-Clifford architectures.
Their doped extensions interpolate toward %
universal quantum computation~\cite{HebenstreitJozsaKrausStrelchuk2019,DiasKoenig2024,ReardonSmithEtAl2024,OszmaniecEtAl2022,PaviglianitiEtAl2026,BallarTriguerosEtAl2026}. 
Sufficiently deep matchgate circuits equilibrate to random fermionic Gaussian states~\cite{BejanBeriMcGinley2025,swann2025spacetime,SierantTurkeshiTarabunga2026,grevink2025glueshortdepthdesignsunitary}, which also arise as eigenstates of quadratic Sachdev--Ye--Kitaev ($\mathrm{SYK}_2$) Hamiltonians~\cite{SachdevYe1993,MaldacenaStanford2016,LydzbaRigolVidmar2020}. Recent studies have shown that, from a Clifford perspective, 
random fermionic Gaussian states are typically close to maximally magical~\cite{Collura2026,BeraSchiro2025,SierantTurkeshiTarabunga2026,Braccia2026,IannottiEtAl2026,ColluraBeriTirrito2026,lastres2026}. Nevertheless, the detailed structure of magic in random fermionic Gaussian states---including the full Pauli spectrum and stabilizer entropies, the role of particle-number symmetry, and the reliability of sampling-based estimates---has remained elusive.

Here we characterize the stabilizer entropies, their filtered counterparts, and the Pauli spectrum of random fermionic Gaussian states, both with and without $\mathrm{U}(1)$ particle-number symmetry. 
For the Haar-random Majorana Gaussian ensemble, we express the ensemble-averaged stabilizer purities at finite $N$ as finite sums of Selberg products and reconstruct the exact ensemble-averaged Pauli spectrum as a mixture of products of independent beta-distributed random variables, resolved by Majorana weight.
For Haar-random Slater determinants at fixed filling, we express the ensemble-averaged stabilizer purity at every positive integer R\'enyi index $q$ as an integral over $\mathrm{SU}(2q)$, whose dimension is independent of $N$. We evaluate the $q=2$ case explicitly as a finite sum and analyze the scaling behavior for $q\geq2$.

We find that the Pauli spectrum of a random Gaussian state is organized into a hierarchy of Majorana-weight sectors: an exponentially large, asymptotically lognormal bulk that appears featureless at accessible sizes, together with rare low-weight tails in which Pauli expectation values are suppressed only polynomially. 
This hierarchy produces a freezing transition at $q_c=2$: the annealed magic density is unity for $1<q\leq2$ and equals $1/(q-1)$ for $q>2$. Haar-random states in the full Hilbert space exhibit the same density, with the $q>2$ branch determined by the identity Pauli string contribution, while removing the identity Pauli string restores unit \emph{filtered} SRE density for every fixed $q>1$~\cite{TurkeshiDymarskySierant2025}.
For random Gaussian states, freezing persists even after Pauli strings corresponding to both identity and parity operators have been filtered.
For this ensemble, both filtered and unfiltered magic densities equal $1/(q-1)$ for each fixed $q>2$. A direct characteristic-sampling estimate based on polynomially many samples instead yields an apparent density tending to $1$. This discrepancy arises because sampling misses the weight-two Majorana strings and their complements that dominate the averaged filtered stabilizer purity: their combined probability under filtered sampling is only $2N/(2^N-2)$.
Subleading corrections distinguish the Majorana, complex-Slater, and real-Slater ensembles and therefore the ground states of the corresponding $\mathrm{SYK}_2$ models, whose nonstabilizerness has been studied numerically in Ref.~\cite{BeraSchiro2025} (see also Ref.~\cite{JasserOdavicHamma2025} for the interacting SYK model).

\section{Overview of the main results}
\label{sec:overview}

In this section, we summarize the main results; the derivations are given in the following sections.

\subsection{Setting}
\label{sec:overview-setting}

The notation and conventions used below are developed in \cref{sec:setup}. We consider $N$ Dirac fermionic modes, equivalently $N$ qubits under the Jordan--Wigner transformation~\cite{JordanWigner1928}, with Hilbert-space dimension $D=2^N$. For a pure state $|\Psi\rangle$, we denote by $x_P=\langle\Psi|P|\Psi\rangle$ the expectation value of a Pauli string $P\in\Pcal_N$. 
The Pauli spectrum is the collection of all $4^N$ such values, or equivalently their distribution $\Pi(x)$, with equal weight assigned to every Pauli string, defined in \cref{eq:pauli-spectrum-def}~\cite{TurkeshiDymarskySierant2025}.
We consider real $q>0$ throughout, with the SRE at $q=1$ defined by continuity. The moments of the nonnegative squared values define the stabilizer purities, $\zeta_q=D^{-1}\sum_P (x_P^2)^q$, and the stabilizer R\'enyi entropies (SREs), $M_q=\log(\zeta_q)/(1-q)$~\cite{LeoneOliveiroHamma2022}. Throughout, unless stated otherwise, $\log\equiv\log_2$, and we use $\ln$ for the natural logarithm.

A pure fermionic Gaussian state is obtained by applying a unitary generated by a quadratic fermion operator to a Fock product state; its explicit many-body definition is given in \cref{eq:gaussian-many-body}. Such a state has definite fermion parity.
Consequently, the identity and total-parity Pauli string $Z^{\otimes N}$ both have unit squared expectation value. Removing these two protected contributions and renormalizing the remainder gives
\begin{equation}
 \widetilde\zeta_q=\frac{D\zeta_q-2}{D-2},
 \qquad
 \widetilde M_q=\frac{\log\widetilde\zeta_q}{1-q},
\end{equation}
as defined in \cref{eq:filtered}. This is the Gaussian-adapted version of the filtered stabilizer entropy introduced in Ref.~\cite{TurkeshiDymarskySierant2025} and agrees with the convention of Ref.~\cite{Collura2026}.

Because we study random-state ensembles, the stabilizer purity fluctuates between states. We distinguish the annealed SRE, obtained by taking the logarithm after averaging the stabilizer purity, from the quenched SRE, obtained by averaging the entropy itself; their definitions are in \cref{eq:ann-que}. For the Majorana ensemble and the half-filled complex Slater ensemble, their difference is at most $O(\log N)$ for each fixed $q\geq2$. This difference is subextensive, so both averages give the same magic density. The precise statement is given in \cref{sec:quenched}.

We analyze three ensembles: the Haar orbit of pairing-allowed Majorana Gaussian states, with covariance matrix $\Gam=OJO^{\mathsf T}$ and $O$ Haar distributed on the special orthogonal group $\mathrm{SO}(2N)$; Haar-random complex Slater determinants at fixed particle number $r$, whose orbit is the complex Grassmannian $\mathrm{Gr}_{\mathbb C}(r,N)$ and whose filling is $\nu=r/N$; and their real counterparts on $\mathrm{Gr}_{\mathbb R}(r,N)$\footnote{The complex Grassmannian is the set of $r$-dimensional subspaces of $\mathbb C^N$, each specifying the occupied one-particle subspace of a Slater determinant. The first $r$ columns of a Haar-random unitary matrix $U\in\mathrm{U}(N)$ span a uniformly sampled subspace. Changing the orthonormal basis within this subspace changes the Slater determinant only by an overall phase. The real case uses subspaces of $\mathbb R^N$ and Haar-random orthogonal matrices in $\mathrm{O}(N)$.}. Under the Jordan--Wigner transformation, every Pauli string corresponds, up to a phase, to a Majorana monomial $\gamma_A$, with $A\subseteq[2N]$. Wick's theorem then expresses every even-weight expectation value as a Pfaffian, so that $x_A^2=\det\Gam_A$, whereas odd-weight strings vanish identically. The Majorana weight $|A|=2k$ therefore decomposes the Pauli spectrum into sectors.

\subsection{The Majorana ensemble: exact moments and exact spectrum}
\label{sec:overview-majorana}

For the Majorana ensemble, rotational invariance reduces all moments to fixed-Majorana-weight sectors. Within each sector, the Jacobi law for the restricted covariance matrix yields a Selberg product. Writing $m=\min(k,N-k)$ and denoting by $\rising{a}{b}=\Gamma(a+b)/\Gamma(a)$ the rising factorial, one obtains
\begin{equation}
\E\bigl[(x_A^2)^q\bigr]
 =\prod_{j=0}^{m-1}\frac{\rising{j+\tfrac12}{q}}{\rising{N-m+j+\tfrac12}{q}},
 \qquad |A|=2k,
\end{equation}
as shown in \cref{eq:R-general}.
Jacobi ensembles of this kind are familiar from fermionic Page-curve calculations~\cite{Bianchi2021,BianchiHacklKieburgEtAl2022}.
Summing over the weight sectors then gives the finite-$N$ annealed stabilizer purity for every real $q>0$, \cref{eq:maj-all-q}. At $q=2$, the sum reduces to the Catalan ratio $\E\zeta_2=2^N/C_{N+1}$, recovering the result of Ref.~\cite{SierantTurkeshiTarabunga2026}.

The sector moments also determine the full averaged Pauli spectrum: because $x_A^2$ lies in $[0,1]$, its probability distribution is uniquely fixed by its integer moments.
Conditional on Majorana weight $2k$, the squared Pauli expectation value is distributed as a product of $m$ independent beta-distributed variables, \cref{eq:beta-product}; summing these conditional laws over the weight sectors gives the exact mixture in \cref{eq:maj-spectrum}. At central weights, $|A|=2k=N+O(\sqrt N)$, the bulk becomes asymptotically lognormal, with $\E[\ln x_A^2]=-N\ln2-\tfrac12\ln N+O(1)$ and variance $\ln N+O(1)$, \cref{eq:central-lognormal}. The complete derivation is given in \cref{sec:majorana}.

\subsection{Freezing of the stabilizer entropies}
\label{sec:overview-freezing}

As developed in \cref{sec:freezing}, the sector decomposition maps the annealed stabilizer purity onto a statistical-mechanics problem. A sector of Majorana weight $2\alpha N$ contributes a combinatorial entropy $2H(\alpha)$ per mode and an energetic cost $qH(\alpha)$ per mode, where $H$ is the binary entropy in bits. Consequently,
\begin{equation}
 \frac1N\log\E\zeta_q
 =-1+\sup_{0\leq\alpha\leq1/2}\,(2-q)H(\alpha)+o(1),
\end{equation}
as stated in \cref{eq:maj-saddle}. For $q<2$, the supremum is attained by the central sector, $\alpha=1/2$; for $q>2$, it moves to the boundary, $\alpha=0$. At the critical index $q_c=2$, all sectors with Majorana weight proportional to $N$ contribute at the same exponential order, with their relative importance determined by subleading factors.
The annealed SREs therefore freeze according to \cref{eq:maj-M-phases}, with unfiltered magic density $D_q=\lim_{N\to\infty}{ M_{q,\mathrm{ann}}}/{N}$ and filtered magic density $\widetilde{D}_q=\lim_{N\to\infty}{\widetilde M_{q,\mathrm{ann}}}/{N}$ coinciding and given by
\begin{equation}
 D_q\equiv
 \widetilde D_q=\begin{cases}
 1, & q\leq2,\\[1mm]
 \dfrac1{q-1}, & q\geq2.
 \end{cases}
\end{equation}

For typical states in the full Hilbert space, the analogous freezing of the unfiltered SRE $D_q^\mathrm{typ}=1/(q-1)$ for $q\ge 2$ is an artifact of the identity string, and filtering restores $\widetilde D_q^\mathrm{typ}=1$ for every fixed $q>1$~\cite{TurkeshiDymarskySierant2025}. The transition in random Gaussian fermions is qualitatively different. Although the filtration removes the identity and parity strings exactly, for every fixed $q>2$ the filtered stabilizer purity scales as $\E\widetilde\zeta_q=\Theta(N^{2-q}2^{-N})$, rather than as the typical-state value $\Theta(2^{-(q-1)N})$. The leading contribution comes not from the pinned strings but from the boundary sectors, beginning with the $\Theta(N^2)$ weight-two Majorana strings and their weight-$(2N-2)$ counterparts, whose squared expectation values are typically of order $N^{-1}$ by Wick's theorem. The freezing moreover survives the removal of any finite collection of weight sectors; see \cref{sec:freezing-highq}.

The boundary sectors also determine the subleading expansion: exactly for the Majorana ensemble, and in the scaling limit for the number-conserving ensembles. With $a_q=\rising{3/2}{q-1}=\Gamma(q+\tfrac12)/\Gamma(\tfrac32)$, the Majorana ensemble gives, for every fixed real $q>2$,
\begin{equation}
 \widetilde M_{q,\mathrm{ann}}
 =\frac{N+(q-2)\log N-\log\bigl(2a_q\bigr)}{q-1}
 +\Delta_q(N),
\end{equation}
which is \cref{eq:maj-filtered-highq}; here $\Delta_q(N)=o(1)$ is the leading algebraic correction to the scaling.

\subsection{\texorpdfstring{{Failure of perfect sampling for $q>2$}}{Failure of perfect sampling for q>2}}
\label{sec:overview-sampling}

After removing identity and parity, the ensemble-averaged $q$th moments of the squared Pauli expectation values are asymptotically dominated by strings with low Majorana weight and their complements for fixed $q>2$.Under characteristic Pauli sampling, a string $A$ is drawn with probability $Y_A/2^N$, where $Y_A=x_A^2$. 
Writing the Majorana weight of the sampled string as $|A|=2K$, we have $K\sim\mathrm{Binomial}(N,1/2)$ for every pure Gaussian state, as proved in \cref{eq:binomial-degree}.
Thus the two sectors that dominate the filtered moments for $q>2$ have combined probability $2N/2^N$ (or $2N/(2^N-2)$ when sampling the filtered distribution), which decays exponentially with the mode number $N$.

Perfect sampling produces independent draws from the correct distribution~\cite{Collura2026}, but resolving a contribution of exponentially small sampling probability still requires exponentially many draws. In \cref{eq:sampling-false-density}, we prove that the direct estimator formed from a polynomial number of samples has limiting entropy density $1$ for every fixed $q>1$. For $q>2$, this differs from the correct density $1/(q-1)$. In particular, the near-unit leading coefficient reported at $q=3$ in Ref.~\cite{Collura2026} is explained by this limitation: the true density tends to $1/2$. Our exact statewise upper bound independently rules out unit asymptotic density for any sequence of pure Gaussian states. The reported scaling is a sampling artifact: direct sampling misses the rare strings that dominate the moment, which require exponentially many draws to resolve.

\subsection{Number-conserving ensembles}
\label{sec:overview-u1}

The calculation for number-conserving Gaussian states is developed in \cref{sec:u1}. For Haar-random Slater determinants with fixed particle number $r$, the $2q$-replica state average is the normalized projector onto a single irreducible $\mathrm{U}(N)$ multiplet. Skew Howe duality~\cite{Howe1989,Panova18skew} then expresses every integer moment as a coefficient integral over $\mathrm{SU}(2q)$, whose dimension depends on $q$ but not on $N$:
\begin{equation}
 \E\zeta_q
 =\frac{1}{d_{N,r}^{(2q)}}\,[z^{2qr}]
 \int_{\mathrm{SU}(2q)}A_q(z,s)^N\dd s.
\end{equation}
This is \cref{eq:u1-all-q}. The local weight $A_q$, defined in \cref{eq:Aq}, combines even elementary-symmetric characters with a signed sum of complementary minors.

At $q=2$, an accidental isomorphism reduces the group integral to $\mathrm{SO}(6)$, yielding the closed finite sum in \cref{eq:u1-q2} (recovering the result in Ref.~\cite{Braccia2026}) and the exact large-deviation rate $I_2(\nu)$ in \cref{eq:I2}.

At half filling, in the scaling limit, the expansion of the complex Slater ensemble in fixed Majorana-weight sectors predicts the same extensive, logarithmic, and constant terms as \cref{eq:maj-filtered-highq}.
The algebraic correction $\Delta_q=o(1)$ is the first term that distinguishes the complex-Slater and Majorana ensembles.

\subsection{The \texorpdfstring{$\mathrm{SYK}_2$}{SYK2} dictionary}
\label{sec:overview-syk}

As detailed in \cref{sec:syk}, the Majorana Gaussian, complex Slater, and real Slater ensembles coincide with the ground-state ensembles of the corresponding $ \mathrm{SYK}_2$ models.
For particle-number-conserving models, complex hopping drawn from the Gaussian unitary ensemble (GUE)~\cite{mehta2004random,Haake2010fgh} at fixed filling generates the complex Slater ensemble; real hopping drawn from the Gaussian orthogonal ensemble (GOE) generates the real-Slater ensemble. At half filling, the real ensemble has $\binom{2N}{N}$ generically nonzero Pauli values, whereas the complex ensemble populates all $2^{2N-1}$ even strings. For the real ensemble, we determine the weight-two moments exactly and obtain the high-$q$ asymptotics by expanding the fixed-weight sectors. This expansion predicts the same extensive frozen coefficient as in the other two ensembles, with different constant and $1/N$ terms; see \cref{sec:syk-discrimination}.

The Pauli spectrum and SREs of $\mathrm{SYK}_2$ were studied using Metropolis sampling in Ref.~\cite{BeraSchiro2025}, while Ref.~\cite{JasserOdavicHamma2025} investigated SREs in the interacting SYK model.
In particular, Ref.~\cite{BeraSchiro2025} numerically reported an approximately exponential Pauli-spectrum bulk, described by $f_N(x)=e^{-|x|/b_N}/(2b_N)$, with $b_N$ fixed by the second-moment normalization.
Our beta-variable analysis motivates this bulk approximation and its triangular semilogarithmic profile, while our low-weight expansion captures the rare-string contributions that control the stabilizer entropies in the frozen regime $q>2$.

\subsection{Annealed versus quenched averages}
\label{sec:overview-quenched}

Because annealed averages can, in principle, be dominated by rare disorder realizations, their relation to typical-state behavior must be established separately. In \cref{sec:quenched}, we prove that such rare realizations do not affect the leading magic density. For the Majorana ensemble and the half-filled complex Slater ensemble, and for every fixed real $q\geq2$, the quenched and annealed stabilizer R\'enyi entropies satisfy
\begin{equation}
 M_{q,\mathrm{que}}
 =M_{q,\mathrm{ann}}+O(\log N)
 =\frac{N}{q-1}+O(\log N).
\end{equation}
This is \cref{eq:quenched-half-density}.

\subsection{Magic under partial trace}
\label{sec:overview-partial-trace}

Finally, in \cref{sec:partial-trace} we ask how much of the system can be discarded before its magic vanishes. 
We witness mixed-state magic using the identity-filtered stabilizer norm $\widetilde{\mathcal D}(\rho)$ defined in \cref{eq:pt-Dnorm-def}. The condition $ \widetilde{\mathcal D}(\rho)>1$ certifies that $\rho$ lies outside the stabilizer polytope~\cite{HowardCampbell2017}.
For the Majorana ensemble, every Pauli expectation value of the reduced state $\rho_n$ after tracing out the rightmost $N-n$ modes is precisely governed by the sector law \cref{eq:R-general}. With retained fraction $c=n/N$ and traced fraction $\kappa=1-c$, both the annealed and typical logarithmic densities of the filtered stabilizer norm equal $2[F(c)-c]$, \cref{eq:pt-annealed-density,eq:pt-Sn-concentration}, where $F(c)$ is the sector-optimized rate \cref{eq:pt-Fdef}. The witness therefore certifies magic for every traced fraction below
\begin{equation}
 \kappa_\star^{\rm G}=0.7612586144\ldots,
\end{equation}
which is \cref{eq:pt-cstar}. 

This threshold lies above the analogous transition of a Haar-random state, whose reduced density matrix enters the stabilizer polytope at $\kappa=2/3$~\cite{LiuLiu2026}. In the resulting window $2/3<\kappa<\kappa_\star^{\rm G}$, the reduced random Gaussian state is still certified to be magical while the Haar-induced state is typically a stabilizer mixture. 
Despite their classical simulability, random Gaussian states therefore exhibit magic that is more resilient to information loss caused by discarding modes than that of Haar-random states, typically regarded as maximally complex.

\section{Pauli spectrum and stabilizer entropies of Gaussian states}
\label{sec:setup}

\subsection{Pauli strings, Majorana strings, and the Pauli spectrum}
\label{sec:setup-pauli}

We consider $N$ qubits with Hilbert-space dimension $D=2^N$. Let $\Pcal_N=\{I,X,Y,Z\}^{\otimes N}$ denote the canonical Hermitian representatives of the $N$-qubit Pauli group modulo overall phases, so that $|\Pcal_N|=4^N$. For a pure state $|\Psi\rangle$, define $x_P=\langle\Psi|P|\Psi\rangle$. The Pauli spectrum is the multiset of these signed expectation values, or equivalently its empirical distribution~\cite{BeverlandCampbellHowardKliuchnikov2020,TurkeshiDymarskySierant2025,dowling2026noiseinducedsimulabilitytransitionoperator,Sarma26uni},
\begin{equation}
 \mathrm{spec}(|\Psi\rangle)
 =\bigl\{x_P=\langle\Psi|P|\Psi\rangle\mid P\in\Pcal_N\bigr\},
 \qquad
 \Pi(x)=\frac{1}{4^N}\sum_{P\in\Pcal_N}\delta(x-x_P).
 \label{eq:pauli-spectrum-def}
\end{equation}
We use the term Pauli spectrum interchangeably for the multiset and the distribution $\Pi(x)$. Since the Pauli strings form an orthogonal operator basis,
$|\Psi\rangle\langle\Psi|=D^{-1}\sum_{P\in\Pcal_N}x_P P$, and purity gives the statewise Parseval identity $\sum_{P\in\Pcal_N}x_P^2=D$. A pure stabilizer state has exactly $D$ nonzero Pauli expectation values, each equal to $\pm1$, while all the remaining ones vanish.

Under the Jordan--Wigner transformation~\cite{JordanWigner1928}, we introduce the $2N$ Hermitian Majorana operators
\begin{equation}
 \gamma_{2i-1}=Z_1\cdots Z_{i-1}X_i,
 \qquad
 \gamma_{2i}=Z_1\cdots Z_{i-1}Y_i,
 \qquad i=1,\ldots,N,
\end{equation}
which satisfy $\{\gamma_\mu,\gamma_\nu\}=2\delta_{\mu\nu}$. For an ordered subset $A=\{\mu_1<\cdots<\mu_{|A|}\}\subseteq[2N]$, with $[2N]=\{1,\ldots,2N\}$, we define the Majorana string
\begin{equation}
 \gamma_A
 =i^{|A|(|A|-1)/2}\gamma_{\mu_1}\cdots\gamma_{\mu_{|A|}}.
 \label{eq:majorana-string}
\end{equation}
This phase convention ensures that $\gamma_A^\dagger=\gamma_A$ and $\gamma_A^2=I$. The $4^N$ operators $\{\gamma_A\}_{A\subseteq[2N]}$ constitute another Hermitian Pauli basis: for every $A$, there is a unique $P_A\in\Pcal_N$ and a sign $s_A\in\{\pm1\}$ such that $\gamma_A=s_A P_A$~\cite{SierantTurkeshiTarabunga2026,TirritoEtAl2025}. We write $x_A=\langle\Psi|\gamma_A|\Psi\rangle$; in particular, $x_A^2=x_{P_A}^2$, as required for the stabilizer moments considered below. We call $|A|$ the \emph{Majorana weight}. It is distinct from the Pauli weight, and, unless stated otherwise, all weight-sector decompositions below refer to the Majorana weight. With the conventions above, the total fermion-parity operator satisfies
\begin{equation}
 Z^{\otimes N}=(-1)^N\gamma_{[2N]}.
\end{equation}

\subsection{Stabilizer entropies and their filtered form}
\label{sec:setup-sre}

For real $q>0$, the stabilizer purity and the corresponding stabilizer R\'enyi entropy (SRE) are defined as~\cite{LeoneOliveiroHamma2022}
\begin{equation}
 \zeta_q=\frac{1}{D}\sum_{P\in\Pcal_N}(x_P^2)^q,
 \qquad
 M_q=\frac{1}{1-q}\log\zeta_q .
 \label{eq:def-Mq}
\end{equation}
Here and throughout, $\log\equiv\log_2$, so all entropies are measured in bits. Parseval's identity ensures that the characteristic distribution $\pi_\Psi(P)=x_P^2/D$ forms a probability distribution. Consequently, $M_q=H_q(\pi_\Psi)-N$, where $H_q$ is the R\'enyi entropy: the SRE is a participation entropy in operator space~\cite{TurkeshiSchiroSierant2023,TurkeshiDymarskySierant2025,2frt-tdg9,HaugPiroliMPS2023,p7xt-s9nz,1tyr-rlbb,ylsz-dm3y,khasseh2026hiddenconformalboundarydata}. Although $\zeta_1=1$ identically, the continuous limit $q\to1$ is nontrivial and gives $M_1=H_1(\pi_\Psi)-N$. For $q\geq2$, the SREs are magic monotones~\cite{LeoneBittel2024,HaugPiroli2023}.

For every pure fermionic Gaussian state considered here, both the identity and total-parity strings are pinned: $x_I^2=x_{Z^{\otimes N}}^2=1$. Indeed, Majorana Gaussian states have definite fermion parity, while a state with fixed particle number $r$ satisfies $\langle Z^{\otimes N}\rangle=(-1)^r$. Following the filtration introduced for typical states in Ref.~\cite{TurkeshiDymarskySierant2025} and adapted to Gaussian states in Ref.~\cite{Collura2026}, we remove these two strings and renormalize the remaining characteristic mass. For $N\geq2$, this gives, state by state,
\begin{equation}
 \widetilde\zeta_q
 =\frac{1}{D-2}\sum_{P\notin\{I,\,Z^{\otimes N}\}}(x_P^2)^q
 =\frac{D\zeta_q-2}{D-2},
 \qquad
 \widetilde M_q=\frac{1}{1-q}\log\widetilde\zeta_q .
 \label{eq:filtered}
\end{equation}
In the limit $q\to1$, the removed strings make no contribution to $M_1$, because $x_P^2\log x_P^2=0$ when $x_P^2=1$, but the change in normalization remains. Hence $\widetilde M_1=\frac{D}{D-2}M_1$. The distinction between \cref{eq:def-Mq,eq:filtered} becomes essential in the high-moment regime, where the pinned strings would otherwise dominate. For Haar-random states in the full Hilbert space, removing the identity gives $\widetilde\zeta_q=\Theta(2^{-(q-1)N})$ and hence $\widetilde M_q/N\to1$ for each fixed $q>1$~\cite{TurkeshiDymarskySierant2025}, giving a maximal scaling of SRE in the leading order. For the Majorana Gaussian ensemble at $q>2$, removing identity and parity instead gives $\E\widetilde\zeta_q=\Theta(N^{2-q}2^{-N})$ and $\widetilde M_{q,\mathrm{ann}}/N\to1/(q-1)$; the derivation is in \cref{sec:freezing}.

For a random-state ensemble, we distinguish between annealed and quenched SREs:
\begin{equation}
 M_{q,\mathrm{ann}}=\frac{1}{1-q}\log\E\zeta_q,
 \qquad
 M_{q,\mathrm{que}}=\frac{1}{1-q}\,\E\log\zeta_q,
 \label{eq:ann-que}
\end{equation}
with analogous definitions for the filtered quantities and continuous extensions at $q=1$. By construction, $M_{q,\mathrm{que}}=\E M_q$, and Jensen's inequality gives $M_{q,\mathrm{ann}}\leq M_{q,\mathrm{que}}$ for $q>1$. The annealed quantities are evaluated in \cref{sec:majorana,sec:u1,sec:syk}, while \cref{sec:quenched} shows that the quenched SREs have the same extensive coefficient as their annealed counterparts.

Finally, the ensemble-averaged Pauli spectrum $\overline\Pi_N(x)=\E\,\Pi(x)$ satisfies
\begin{equation}
 D\int (x^2)^q\,\overline\Pi_N(x)\dd x=\E\zeta_q .
 \label{eq:rho-moment}
\end{equation}
Since $x^2\in[0,1]$, the nonnegative integer moments in \cref{eq:rho-moment} uniquely determine the ensemble-averaged distribution of the squared Pauli expectation values, or equivalently of their magnitudes. Recovering the full signed spectrum requires, in addition, the distribution of signs; for the Majorana ensemble it follows from Haar invariance, see \cref{eq:signed-spectrum-complete}.

\subsection{Gaussian states and random ensembles}
\label{sec:setup-ensembles}

We define the Dirac fermion annihilation operators $c_i=(\gamma_{2i-1}+i\gamma_{2i})/2$, with $c_i|0\rangle=0$. The Fock product states are
\begin{equation}
 |\mathbf n\rangle=(c_1^\dagger)^{n_1}\cdots(c_N^\dagger)^{n_N}|0\rangle,
 \qquad n_i\in\{0,1\}.
 \label{eq:fock-seed}
\end{equation}
A pure fermionic Gaussian many-body state is of the form~\cite{Bravyi2005,SuraceTagliacozzo2022}
\begin{equation}
 |\Psi_G\rangle=U_K|\mathbf n\rangle,
 \qquad U_K=\exp\!\left(\frac14\sum_{\mu,\nu=1}^{2N}K_{\mu\nu}\gamma_\mu\gamma_\nu\right),
 \qquad K=-K^{\mathsf T}\in\mathbb R^{2N\times2N}.
 \label{eq:gaussian-many-body}
\end{equation}
For real antisymmetric $K$, $U_K$ is unitary and rotates the Majorana operators linearly, $U_K^\dagger\boldsymbol\gamma U_K=e^K\boldsymbol\gamma$. It preserves the parity $(-1)^{\sum_i n_i}$ of the reference state. Equivalently, $|\Psi_G\rangle$ is the vacuum of canonical Bogoliubov modes,
\begin{equation}
 \beta_i=\sum_j(u_{ij}c_j+v_{ij}c_j^\dagger),
 \qquad \beta_i|\Psi_G\rangle=0,
 \qquad\{\beta_i,\beta_j^\dagger\}=\delta_{ij},\quad\{\beta_i,\beta_j\}=0.
 \label{eq:bogoliubov-vacuum}
\end{equation}
This definition includes both Slater determinants and pairing-allowed states; we consider pure states throughout.

Up to an overall phase, a pure fermionic Gaussian state is completely characterized by its covariance matrix $\Gam_{\mu\nu}=-\tfrac{i}{2}\langle[\gamma_\mu,\gamma_\nu]\rangle$, a real antisymmetric matrix satisfying $\Gam^2=-\id_{2N}$~\cite{Bravyi2005,SuraceTagliacozzo2022}. With the phase convention of \cref{eq:majorana-string}, Wick's theorem gives $x_A=(-1)^k\Pf(\Gam_A)$ for $|A|=2k$, whereas strings of odd Majorana weight vanish identically. Hence
\begin{equation}
 x_A^2=\Pf(\Gam_A)^2=\det\Gam_A\in[0,1],
 \label{eq:principal-minor}
\end{equation}
where $\Gam_A$ denotes the principal submatrix indexed by $A$. For a pure covariance matrix, Pfaffian complementation yields $x_A^2=x_{A^c}^2$; hence the squared spectrum is symmetric under $k\leftrightarrow N-k$. We write $Y_A=x_A^2$ when convenient.

We consider the following three ensembles.
\begin{enumerate}[label=(\roman*)]
 \item \emph{Majorana Haar ensemble.} Let $\Gam=OJO^{\mathsf T}$, where $O$ is Haar distributed on $\mathrm{SO}(2N)$ and $J=\bigoplus_{j=1}^N\bigl(\begin{smallmatrix}0&1\\-1&0\end{smallmatrix}\bigr)$. 
The induced distribution of $\Gam$ is the invariant measure on the fixed-parity manifold of pure pairing-allowed (Bogoliubov--de Gennes) Gaussian states, namely the coset $\mathrm{SO}(2N)/\mathrm{U}(N)$~\cite{SierantTurkeshiTarabunga2026}. Passing to the parity-flipped component of $\mathrm{O}(2N)$ leaves all statistics of $Y_A$, and therefore all stabilizer entropies, unchanged.

 \item \emph{Complex Slater ensemble.} At fixed particle number $r$ in $N$ modes, let
 \begin{equation}
  |\psi_U\rangle=\prod_{j=1}^{r}\biggl(\sum_{i=1}^{N}U_{ij}\,c_i^\dagger\biggr)|0\rangle,
  \qquad U\sim\mathrm{Haar}[\mathrm{U}(N)].
  \label{eq:slater-ensemble}
 \end{equation}
 The product occupies $r$ orthonormal one-particle orbitals. 
The resulting Slater states are distributed according to the invariant measure on the complex Grassmannian $\mathrm{Gr}_{\mathbb C}(r,N)$ of rank-$r$ correlation projectors
 \begin{equation}
  G=U\diag(\id_r,0)U^\dagger,
  \qquad G_{ij}=\langle c_j^\dagger c_i\rangle.
 \end{equation}
 We denote the filling by $\nu=r/N$.

 \item \emph{Real Slater ensemble.} Restricting the orbital rotation in (ii) to $O\sim\mathrm{Haar}[\mathrm{O}(N)]$ gives the invariant measure on the real Grassmannian $\mathrm{Gr}_{\mathbb R}(r,N)$.
\end{enumerate}
The correspondence between these ensembles and the ground states of quadratic $\mathrm{SYK}_2$ Hamiltonians is established in \cref{sec:syk}. For the number-conserving ensembles (ii)--(iii), let $\mathcal R(G)$ denote the realification of the one-particle correlation matrix $G$, whose $(i,j)$ block in the interleaved Majorana basis is
\begin{equation}
 [\mathcal R(G)]_{ij}=
 \begin{pmatrix}
  \Re G_{ij} & -\Im G_{ij}\\
  \Im G_{ij} & \Re G_{ij}
 \end{pmatrix}.
\end{equation}
Their Majorana covariance matrix is then $\Gam=J[\id_{2N}-2\mathcal R(G)]$, so \cref{eq:principal-minor} applies to all three ensembles. For real $G$, this reduces to $\mathcal R(G)=G\otimes\id_2$.

In the Majorana ensemble, rotations map any set of $2k$ Majoranas to any other set of the same weight. All $\binom{2N}{2k}$ strings of weight $|A|=2k$ therefore have the same probability distribution of $Y_A$ and the same moment $\E[Y_A^q]$ at every finite $N$. We call this collection the weight-$2k$ sector. For number-conserving ensembles we still group strings by their Majorana weight, but strings within the same sector can have different distributions because the symmetry group is smaller. The Majorana ensemble is analyzed in \cref{sec:majorana,sec:freezing}, and the number-conserving ensembles in \cref{sec:u1,sec:syk}.

\section{The Majorana ensemble: exact moments and exact spectrum}
\label{sec:majorana}

\subsection{The Selberg sector law and all moments}
\label{sec:majorana-selberg}

Choose a set $A$ of $2k$ Majorana operators, with $1\leq k\leq N/2$, and consider the corresponding covariance submatrix $\Gam_A$. Its eigenvalues occur in pairs $\pm i\sqrt{\xi_j}$, where $\xi_j\in[0,1]$. The squared singular values $\xi_1,\ldots,\xi_k$, counted once per pair, have joint probability density
\begin{equation}
   p(\xi_1,\ldots,\xi_k)\propto
   \prod_{a<b}(\xi_a-\xi_b)^2
   \prod_{a=1}^{k}\xi_a^{-1/2}(1-\xi_a)^{N-2k}.
   \label{eq:app-jacobi}
\end{equation}
This distribution is a Jacobi ensemble, also encountered in fermionic Page-curve calculations~\cite{Bianchi2021}. Since $x_A^2=\det\Gam_A=\prod_{j=1}^{k}\xi_j$, the moments of the squared Pauli expectation value can be evaluated by integrating against this density.
Combining the corresponding Selberg integral with Pfaffian complementation extends the result to every Majorana-weight sector; the derivation is given in \cref{app:selberg}. 
By rotational invariance, the probability distribution of the squared Pauli expectation value $x_A^2$ depends on a subset $A$ only through its weight $|A|=2k$. Setting $m=\min(k,N-k)$ and recalling that $Y_A=x_A^2$, we obtain, for every real $q>0$,
\begin{equation}
 R_{N,k}(q):=\E\bigl[Y_A^{\,q}\bigr]
 =\prod_{j=0}^{m-1}
 \frac{\rising{j+\tfrac12}{q}}{\rising{N-m+j+\tfrac12}{q}} .
 \label{eq:R-general}
\end{equation}
Here and below, an empty product is understood as unity, so the endpoint sectors $k=0,N$ give $R_{N,0}(q)=R_{N,N}(q)=1$. For positive integer $q$, the equivalent representation
\begin{equation}
 R_{N,k}(q)=\prod_{a=0}^{q-1}
 \frac{\rising{a+\tfrac12}{k}\rising{a+\tfrac12}{N-k}}{\rising{a+\tfrac12}{N}}
 \label{eq:R-integer}
\end{equation}
makes the complement symmetry $k\leftrightarrow N-k$ manifest.

For $q>0$, all odd-weight strings contribute zero to the stabilizer purity. Summing the even-weight sectors with multiplicity $\binom{2N}{2k}$ therefore gives the exact finite-$N$ annealed stabilizer purity
\begin{equation}
 \begin{split}
 \E\zeta_q^{\mathrm{Maj}}
 &=2^{-N}\sum_{k=0}^{N}\binom{2N}{2k}R_{N,k}(q),\qquad q>0,\\[1mm]
 \E\zeta_q^{\mathrm{Maj}}
 &=2^{-N}\;{}_qF_{q-1}\!\left(
 \begin{matrix}-N,\ \tfrac32,\ \tfrac52,\ldots,\ q-\tfrac12\\
 -N-\tfrac12,\ -N-\tfrac32,\ldots,\ \tfrac32-N-q\end{matrix}
 \;\middle|\;(-1)^q\right),\quad q\in\mathbb N_{>0}.
 \end{split}
 \label{eq:maj-all-q}
\end{equation}
Here, ${}_qF_{q-1}$ denotes the generalized hypergeometric function.
The second line holds for positive integer $q$; the parameter lists after $-N$ are empty at $q=1$. The upper parameter $-N$ terminates the series at degree $N$, and the derivation is given in \cref{app:selberg}.

For $N\geq2$, removing the identity and total-parity sectors, corresponding respectively to $k=0$ and $k=N$, gives the exact filtered moment
\begin{equation}
 \E\widetilde\zeta_q^{\mathrm{Maj}}
 =\frac{\sum_{k=1}^{N-1}\binom{2N}{2k}R_{N,k}(q)}{2^N-2}
 =\frac{2^N\,\E\zeta_q^{\mathrm{Maj}}-2}{2^N-2}.
 \label{eq:maj-filtered-exact}
\end{equation}

\subsection{The second moment}
\label{sec:majorana-q2}

At $q=2$, the finite sector sum in \cref{eq:maj-all-q} can be evaluated in closed form:
\begin{equation}
 \E\zeta_2^{\mathrm{Maj}}
 =\frac{2^N}{C_{N+1}}
 =\frac{2^{N-1}\,N!\,(N+2)!}{(2N+1)!}.
 \label{eq:maj-q2}
\end{equation}
Here $C_n=(n+1)^{-1}\binom{2n}{n}$ is the $n$th Catalan number, and \cref{eq:maj-q2} recovers the result of Ref.~\cite{SierantTurkeshiTarabunga2026}.

Since $M_{2,\mathrm{ann}}=-\log\E\zeta_2$, the Stirling expansion of \cref{eq:maj-q2} gives
\begin{align}
 M_{2,\mathrm{ann}}^{\mathrm{Maj}}
 &=N-\frac32\log N+\log\frac{4}{\sqrt\pi}
 -\frac{21\loge}{8N}+\frac{19\loge}{8N^2}
 +O(N^{-3}),
 \label{eq:maj-q2-asymp}\\
 \widetilde M_{2,\mathrm{ann}}^{\mathrm{Maj}}
 &=N-\frac32\log N+\log\frac{4}{\sqrt\pi}
 -\frac{21\loge}{8N}
 +\frac{8\loge}{\sqrt\pi\,N^{3/2}}
 +\frac{19\loge}{8N^2}
 +O(N^{-5/2}).
 \label{eq:maj-q2-filtered-asymp}
\end{align}
The second line follows from the exact filtration identity \cref{eq:maj-filtered-exact}. Thus, through order $N^{-2}$, the only effect of filtration is the additional positive term $8\loge/(\sqrt\pi N^{3/2})$, generated by removing the identity and parity strings.

For comparison, a generic Haar-random state has $\widetilde M_2^{\mathrm{typ}}=N-\log3+o(1)$~\cite{TurkeshiDymarskySierant2025}. The Majorana ensemble therefore exhibits a magic deficit of $\frac32\log N+O(1)$ already at the critical index $q_c=2$. This logarithmic deficit is the first sign of the freezing analyzed in \cref{sec:freezing}: at $q=2$, all Majorana-weight sectors are marginal at exponential order, whereas for $q>2$ the filtered stabilizer purity is dominated by the low-weight boundary sectors.

\subsection{The exact averaged Pauli spectrum}
\label{sec:majorana-spectrum}

The Gamma-function ratio in \cref{eq:R-general} is the Mellin transform of a product of independent beta variables; see \cref{app:spectrum}. Writing $Y_{N,k}$ for a representative squared expectation value in the sector $|A|=2k$, and setting $m=\min(k,N-k)$, we have
\begin{equation}
 Y_{N,k}\overset{\mathrm d}{=}\prod_{j=0}^{m-1}B_j,
 \qquad
 B_j\sim\operatorname{Beta}\Bigl(j+\frac12,\;N-m\Bigr),
 \label{eq:beta-product}
\end{equation}
where the factors are mutually independent and $\overset{\mathrm d}{=}$ denotes the equality of probability distributions.
For $m=0$, the product is empty and $Y_{N,k}=1$ deterministically.

To relate the squared spectrum to \cref{eq:pauli-spectrum-def}, choose one of the $4^N$ Pauli strings uniformly and record $Y=x_P^2$. Its ensemble-averaged probability distribution is
\begin{equation}
 \overline\rho_N(y)
 =\E\!\left[4^{-N}\sum_{P\in\Pcal_N}\delta(y-x_P^2)\right]
 =\int_{-1}^{1}\overline\Pi_N(x)\,\delta(y-x^2)\dd x.
 \label{eq:rho-pi-relation}
\end{equation}
Thus $\overline\rho_N$ counts squared values with uniform weight per string. Half of the strings have odd Majorana weight and give $Y=0$. Combining the remaining weight sectors gives
\begin{equation}
 \overline\rho_N
 =\frac12\,\delta_0
 +4^{-N}\sum_{k=0}^{N}\binom{2N}{2k}\Law\bigl(Y_{N,k}\bigr).
 \label{eq:maj-spectrum}
\end{equation}
Here $\Law(Y_{N,k})$ denotes the probability distribution of the beta product $Y_{N,k}$ in \cref{eq:beta-product}. The notation $\delta_0$ denotes a delta peak at $Y=0$: its coefficient $1/2$ is the fraction of identically vanishing odd-weight strings. Identity and parity give two further delta peaks at $Y=1$, each with weight $4^{-N}$. The other even-weight sectors have continuous distributions on $0<Y<1$. Their probabilities, together with the delta peaks, sum to one.

To recover the signed canonical Pauli spectrum, we write the density of the continuous part of the distribution in \cref{eq:maj-spectrum} as
\begin{equation}
 \overline\rho_N^\circ(y)
 =4^{-N}\sum_{k=1}^{N-1}\binom{2N}{2k}f_{N,\min(k,N-k)}(y),
 \label{eq:continuous-squared-spectrum}
\end{equation}
where $f_{N,m}$ is the density of the beta product. The total mass of $\overline\rho_N^\circ$ is $(2^{2N-1}-2)/4^N$: the number of nontrivial even strings divided by the total number of Pauli strings. In every nontrivial even sector, Haar invariance gives equal probabilities for positive and negative expectation values. Splitting $Y$ equally between $x=\sqrt Y$ and $x=-\sqrt Y$, and using $\dd Y=2|x|\dd x$, gives the full signed spectrum
\begin{equation}
 \overline\Pi_N(x)
 =\frac12\delta(x)+4^{-N}\bigl[\delta(x-1)+\delta(x-p)\bigr]
   +|x|\,\overline\rho_N^\circ(x^2),
 \label{eq:signed-spectrum-complete}
\end{equation}
where the last term is supported on $0<|x|<1$, and $p=\langle Z^{\otimes N}\rangle=\pm1$ is the fermion parity. The identity operator always gives $x=1$; the parity Pauli string gives $x=p$. For the orbit generated from the vacuum, $p=+1$, so both endpoint contributions sit at $x=1$. In the odd-parity component, one sits at each of $x=\pm1$.

For $m\geq1$, each sector has the continuous Meijer $G$-function density given in \cref{eq:meijer-density}. The logarithmic spectrum has a simpler description: conditional on Majorana weight $2k$, the cumulants of $\ln Y$ are finite sums of polygamma functions, as shown in \cref{eq:log-cumulants}. In the binomially dominant central window $k=N/2+O(\sqrt N)$,
\begin{equation}
 \E[\ln Y]=-N\ln2-\frac12\ln N+O(1),
 \qquad
 \Var(\ln Y)=\ln N+O(1),
 \label{eq:central-lognormal}
\end{equation}
while every fixed higher cumulant is $O(1)$. In other words, $Y$ is asymptotically lognormal in the central fluctuation window. For comparison, a nonidentity Pauli value of a Haar-random state in the full Hilbert space has $2^NY$ asymptotically distributed according to the one-degree-of-freedom chi-square (Porter--Thomas) law~\cite{TurkeshiDymarskySierant2025,PorterThomas1956}. The low-weight tails of the Gaussian-state spectrum, outside this central window, dominate the moments for $q>2$; see \cref{sec:freezing}.

\begin{figure}[t]
 \centering
 \includegraphics[width=0.7\linewidth]{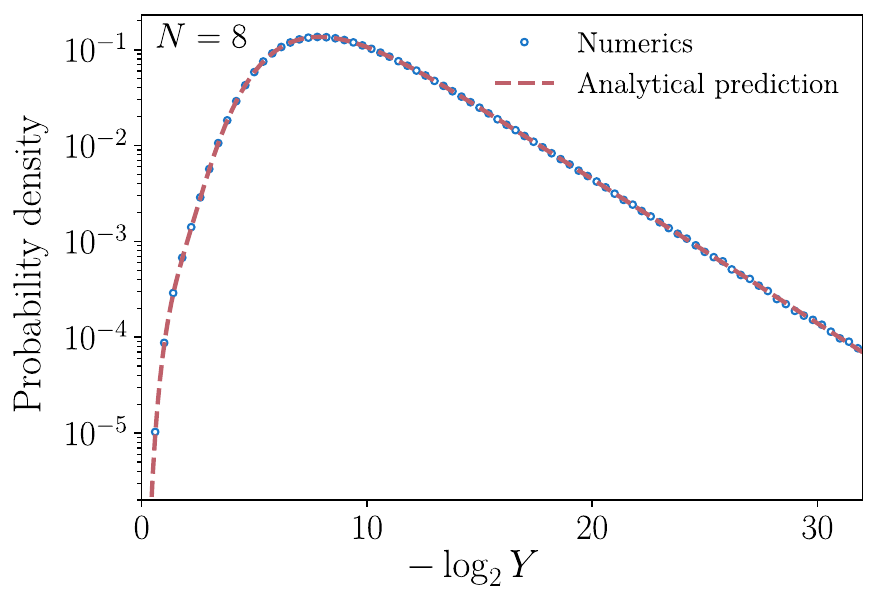}
 \caption{Squared Pauli spectrum of the Majorana ensemble at $N=8$, conditioned on nontrivial even strings and plotted as $z=-\log_2Y$. Symbols: every such Pauli expectation evaluated from the covariance matrix in each of $1024$ independently drawn states, using Wick's theorem. Dashed curve: the analytical density \cref{eq:plotted-log-spectrum}, with the sector density from \cref{eq:meijer-density}.}
 \label{fig:spectrum}
\end{figure}

In \cref{fig:spectrum}, we retain only the $2^{2N-1}-2$ nontrivial even strings and normalize their distribution to one. The plotted density of $z=-\log_2Y$ is therefore
\begin{equation}
 g_N(z)=\frac{(\ln2)\,2^{-z}}{2^{2N-1}-2}
 \sum_{k=1}^{N-1}\binom{2N}{2k}
 f_{N,\min(k,N-k)}(2^{-z}),\qquad z>0.
 \label{eq:plotted-log-spectrum}
\end{equation}
The curve follows from \cref{eq:beta-product,eq:meijer-density}. For the symbols, we draw independent Haar-distributed covariance matrices, corresponding to Haar-random Gaussian states, and evaluate every even Pauli expectation as a Pfaffian using Wick's theorem. %

\section{Freezing of the annealed stabilizer entropies}
\label{sec:freezing}

\subsection{Sector thermodynamics and the transition at
\texorpdfstring{$q_c=2$}{qc=2}}
\label{sec:freezing-saddle}

The exact moments derived in \cref{sec:majorana} undergo a sharp change of behavior as the R\'enyi index is increased. The origin of the transition is best seen in the sector representation \cref{eq:maj-all-q}, which may be viewed as a partition function over Majorana weights~\cite{Tarabunga2024,Salazar2026}: the exponentially large multiplicity of a sector competes with the magnitude of the Pauli expectation values within it. Let $k=\alpha N$, with $0\leq\alpha\leq1/2$, and denote by $H(\alpha)=-\alpha\log\alpha-(1-\alpha)\log(1-\alpha)$ the binary entropy in bits. For fixed $q>1$, Stirling asymptotics of the sector multiplicity and the Selberg ratio give, uniformly for $\alpha$ in compact subsets of $(0,1/2]$,
\begin{equation}
 \frac1N\log\binom{2N}{2\alpha N}=2H(\alpha)+o(1),
 \qquad
 \frac1N\log R_{N,\alpha N}(q)=-qH(\alpha)+o(1).
 \label{eq:maj-sector-rate}
\end{equation}
Together with the fixed-weight boundary analysis below, these exponential rates yield
\begin{equation}
 \frac1N\log\E\zeta_q^{\mathrm{Maj}}
 =-1+\sup_{0\leq\alpha\leq1/2}(2-q)\,H(\alpha)+o(1).
 \label{eq:maj-saddle}
\end{equation}
For $1<q<2$, the supremum is attained by the central sector $\alpha=1/2$. For $q>2$, the exponent instead favors the boundary $\alpha=0$, whose dominance is established by the fixed-weight expansion \cref{eq:fixed-degree-tower}. At $q=2$, all macroscopic weights have the same exponential rate and are therefore marginal at exponential order. Consequently,
\begin{equation}
 \lim_{N\to\infty}\frac1N\log\E\zeta_q^{\mathrm{Maj}}
 =\begin{cases}
 1-q, & 1<q<2,\\
 -1, & q\geq2,
 \end{cases}
 \label{eq:maj-free-energy-limit}
\end{equation}
so the annealed free-energy density freezes at the critical index $q_c=2$. This mechanism is analogous to the freezing of the so-called participation entropies~\cite{LuitzAletLaflorencie2014,Mace2019,SierantTurkeshi2022,Liu25ipr,TurkeshiSierant2024,tirrito2025universalspreadingnonstabilizernessquantum,1jzy-sk9r,lkwg-4dbt,PhysRevLett.134.010401,aditya2026coherencedynamicsquantummanybody,heinrich2026criticalbehaviorsmagicparticipation,liu2026diffusiverelaxationparticipationentropy,xiao2026diffusivedynamicsnonstabilizerness,Magni2025quantumcomplexity,p8dn-glcw,w6l4-mh6k} in multifractal wave functions and in the random-energy model~\cite{Derrida1981,EversMirlin2008}.

Beyond the leading exponential rate, the three regimes are distinguished by their finite-size scaling:
\begin{equation}
 \E \zeta_q^{\mathrm{Maj}}\sim
 \begin{cases}
 B(q)\,N^{q(q-1)/2}\,2^{(1-q)N}, & 1<q<2,\\[1mm]
 2^N/C_{N+1}, & q=2,\\[1mm]
 2^{1-N}\bigl[1+a_q\,N^{2-q}+o(N^{2-q})\bigr], & q>2,
 \end{cases}
 \label{eq:maj-phases}
\end{equation}
where $B(q)>0$ is independent of $N$ and $a_q=\rising{3/2}{q-1}$. The corresponding annealed SREs therefore obey
\begin{equation}
 M_{q,\mathrm{ann}}^{\mathrm{Maj}}=
 \begin{cases}
 N-\dfrac q2\log N+O(1), & 1<q<2,\\[2mm]
 N-\dfrac32\log N+O(1), & q=2,\\[2mm]
 \dfrac{N-1}{q-1}+o(1), & q>2.
 \end{cases}
 \label{eq:maj-M-phases}
\end{equation}
These expansions are not uniform as $q\to2$. Filtering changes the subleading terms, crucially so for $q>2$, but leaves the thermodynamic magic density unchanged; the filtered asymptotics are derived in \cref{sec:freezing-highq}.

For $q>2$, the maximum at $\alpha=0$ means that the relevant Majorana weight stays finite while $N$ grows. By complementation, weights whose distance from $2N$ stays finite contribute equally. We refer to these two ends of the weight range as boundary sectors. To determine their contributions, we hold the integer $k$ fixed rather than the ratio $k/N$. The weight-$2k$ contribution to $D\,\E\zeta_q^{\mathrm{Maj}}$ is, for every fixed $k\geq0$ and $q>1$,
\begin{equation}
 \binom{2N}{2k}R_{N,k}(q)
 \sim\frac{2^{2k}}{(2k)!}
 \Bigl[\prod_{j=0}^{k-1}\rising{j+\tfrac12}{q}\Bigr]\,
 N^{(2-q)k}.
 \label{eq:fixed-degree-tower}
\end{equation}
The sector $k=N-k_0$ has the same contribution as $k=k_0$. At $q=2$, every fixed-$k$ term is $O(1)$ in system size. At $q>2$, increasing $k$ adds a factor $N^{2-q}$ and hence suppresses the contribution. After removing $k=0,N$, the leading terms are therefore the weight-two sector $k=1$ and the weight-$(2N-2)$ sector $k=N-1$.

\begin{figure}[t]
 \centering
 \includegraphics[width=0.7\linewidth]{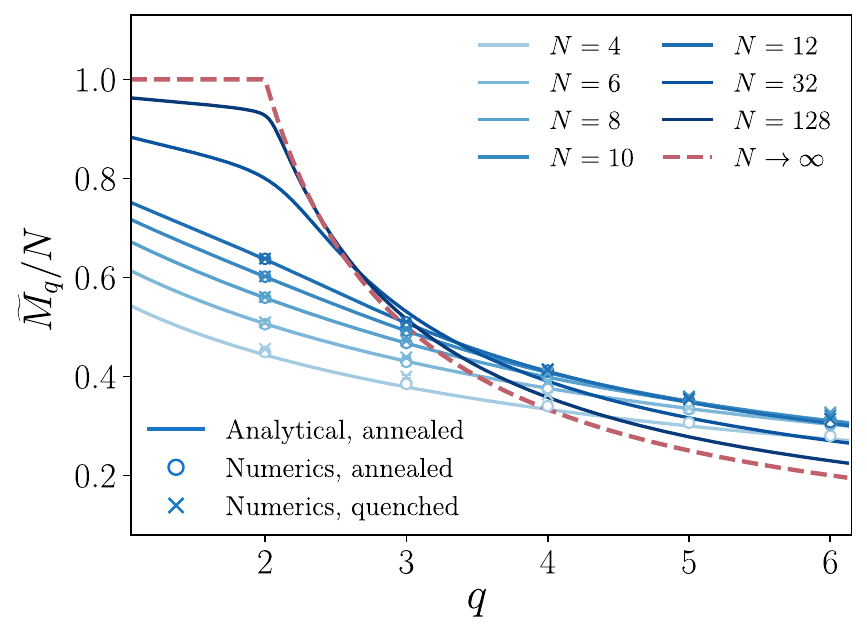}
 \caption{Filtered magic density in the Majorana ensemble. Solid curves: exact annealed result \cref{eq:maj-filtered-exact} at $N=4,6,8,10,12,32,128$. Circles and crosses: annealed and quenched averages at $N=4,6,8,10,12$, respectively, calculated from all nontrivial even strings at integer $q$. Error bars are standard errors across states, propagated through the logarithm for the annealed values. The larger sizes show the approach to the dashed thermodynamic limit \cref{eq:filtered-density}.}
 \label{fig:maj-freezing}
\end{figure}

\Cref{fig:maj-freezing} shows the annealed and quenched numerical averages from $N=4$ to $12$, together with the exact annealed curves extended to $N=32,128$. The quenched values lie above the annealed ones, as required by Jensen's inequality. \Cref{sec:quenched} proves that their magic densities coincide in the thermodynamic limit.

\subsection{High-\texorpdfstring{$q$}{q} asymptotics of the filtered SRE}
\label{sec:freezing-highq}

For $q>2$, the unfiltered stabilizer purity in \cref{eq:maj-phases} is asymptotically saturated by the pinned identity and parity strings. Filtering removes these two contributions and leaves the weight-two strings and their complements as the leading terms. For every fixed real $q>2$, the exact Majorana sector sum gives
\begin{equation}
 \widetilde M_{q,\mathrm{ann}}^{\mathrm{Maj}}
 =\frac{N+(q-2)\log N-\log(2a_q)}{q-1}
 +\Delta_q^{\mathrm{Maj}}(N)+\ldots,
 \label{eq:maj-filtered-highq}
\end{equation}
where $a_q=\rising{3/2}{q-1}$ and the first algebraic correction is
\begin{equation}
 \Delta_q^{\mathrm{Maj}}(N)=
 \begin{cases}
 -\dfrac{d_q\loge}{q-1}\,N^{2-q}, & 2<q<3,\\[2mm]
 -\dfrac{19\loge}{16\,N}, & q=3,\\[2mm]
 +\dfrac{(q-1)\loge}{2\,N}, & q>3,
 \end{cases}
 \qquad
 d_q=\frac{a_q(q+\tfrac12)}{3}.
 \label{eq:maj-delta}
\end{equation}
The three regimes differ in the origin of the leading correction: for $2<q<3$, it comes from the weight-four sector; for $q>3$, from the finite-$N$ expansion of the leading weight-two sector. At $q=3$, both contributions enter at order $N^{-1}$ and combine to give
\begin{equation}
 \widetilde M_{3,\mathrm{ann}}^{\mathrm{Maj}}
 =\frac{N+\log N-\log(15/2)}{2}
 -\frac{19\loge}{16N}+O(N^{-2}).
 \label{eq:maj-q3-filtered}
\end{equation}
Together with the $q\leq2$ branch in \cref{eq:maj-M-phases}, this yields, for $q>1$, the annealed filtered magic density
\begin{equation}
 \widetilde D_q
 =\lim_{N\to\infty}\frac{\widetilde M_{q,\mathrm{ann}}}{N}
 =\min\Bigl\{1,\frac{1}{q-1}\Bigr\},
 \qquad q>1,
 \label{eq:filtered-density}
\end{equation}
whose $q>2$ branch is frozen. For each fixed $q>2$, the algebraic correction $\Delta_q^{\mathrm{Maj}}(N)$ tends to zero. It therefore changes neither the extensive term, the logarithmic term, nor the constant already displayed in \cref{eq:maj-filtered-highq}; it only refines the approach to those terms.

The Gaussian freezing survives the removal of identity and parity because the expectation values of nontrivial low-weight strings remain much larger than those of a typical Haar-random state. For a Haar-random state, the identity alone causes the apparent freezing of the unfiltered SRE; once it is removed, the filtered stabilizer purity scales as $\widetilde\zeta_q^{\,\mathrm{typ}}=\Theta(2^{-(q-1)N})$, and hence $\widetilde D_q^{\mathrm{typ}}=1$ for every fixed $q>1$~\cite{TurkeshiDymarskySierant2025}. In the Gaussian case, instead, the filtration in \cref{eq:filtered} removes both identity and parity exactly, leaving the hierarchy of Majorana-weight sectors generated by Wick's theorem. There are $\binom{2N}{2}=N(2N-1)=\Theta(N^2)$ weight-two strings, and their statewise sector average obeys
\begin{equation}
 \frac{1}{\binom{2N}{2}}\sum_{a<b}x_{\{a,b\}}^2
 =\frac{1}{\binom{2N}{2}}\sum_{a<b}\Gam_{ab}^2
 =\frac{1}{2N-1}.
\end{equation}
Thus the characteristic scale of a squared weight-two expectation value is $x_A^2=\Gam_{ab}^2=O(N^{-1})$: it is algebraically, rather than exponentially, small. The $\Theta(N^2)$ such strings then give
\begin{equation}
 \E\widetilde\zeta_q^{\,\mathrm{Maj}}
 =\Theta\bigl(N^{2-q}\,2^{-N}\bigr)
 \qquad\text{against}\qquad
 \widetilde\zeta_q^{\,\mathrm{typ}}
 =\Theta\bigl(2^{-(q-1)N}\bigr).
 \label{eq:poly-vs-exp}
\end{equation}
Both stabilizer purities are exponentially small in $N$, but their exponential rates differ: the stabilizer purity of Gaussian states has the frozen factor $2^{-N}$, dressed by the algebraic prefactor $N^{2-q}$, whereas the typical-state stabilizer purity scales as $2^{-(q-1)N}$. Their ratio is $\Theta\!\left(N^{2-q}2^{(q-2)N}\right)$ and is therefore exponentially large for every fixed $q>2$.

The frozen density also survives removal of any fixed number of the lowest-weight sectors and their complements. %
If the entire weight-two sector and its complement are discarded, the weight-four sector and its complement become the leading boundary contribution, scaling as $N^{2(2-q)}$ in the unnormalized filtered power sum. More generally, removing any finite number of the lowest-weight sectors and their complements makes the next fixed-weight sector in \cref{eq:fixed-degree-tower} dominant. This changes the logarithmic finite-size term but leaves the frozen density $\widetilde D_q=1/(q-1)$ unchanged.

\subsection{\texorpdfstring{Failure of perfect sampling for $q>2$}{Failure of perfect sampling for q>2}}
\label{sec:freezing-sampling}

With perfect sampling, each Pauli string is drawn exactly from the target distribution and independently of previous draws, so neither equilibration bias nor autocorrelations arise~\cite{Collura2026}.
The accuracy of a moment estimate also depends on how often the strings carrying that moment are drawn. We show that, for the Majorana ensemble, the direct estimator of a filtered SRE with $q>2$ fails with a polynomial sampling budget even when a perfect sampler generates independent strings with exactly the prescribed probabilities.

For a fixed pure Gaussian state $|\Psi\rangle$, let $A_1,\ldots,A_M$ be independent samples from the filtered characteristic distribution $\widetilde\pi_\Psi$ defined in \cref{eq:direct-sampling-estimator}.
The filtered characteristic distribution and the direct stabilizer purity and entropy estimators are
\begin{equation}
 \begin{split}
 \widetilde\pi_\Psi(A)&=\frac{Y_A}{D-2},\qquad A\ne\varnothing,[2N],\\
 \widehat{\widetilde\zeta}_q&=\frac1M\sum_{j=1}^{M}Y_{A_j}^{q-1},
 \qquad
 \widehat{\widetilde M}_q=\frac{\log\widehat{\widetilde\zeta}_q}{1-q}.
 \end{split}
 \label{eq:direct-sampling-estimator}
\end{equation}
For each state, $\E_{\mathrm{samp}}\widehat{\widetilde\zeta}_q=\widetilde\zeta_q$: the stabilizer purity estimator is unbiased. This does not ensure small relative error or an accurate estimate of the filtered SRE.
Its exact relative variance is~\cite{TarabungaTirritoChandaDalmonte2023}
\begin{equation}
 \frac{\Var_{\mathrm{samp}}(\widehat{\widetilde\zeta}_q)}{\widetilde\zeta_q^{\,2}}
 =\frac1M\left(\frac{\widetilde\zeta_{2q-1}}{\widetilde\zeta_q^{\,2}}-1\right),
 \qquad q>1.
 \label{eq:direct-sampling-variance}
\end{equation}

The sampling mass of each Majorana-weight sector can be obtained without averaging over states. 
Expanding $\det(\id+t\Gam)$ in principal minors and using $\Gam^2=-\id$, we obtain
\begin{equation}
 \sum_{A\subseteq[2N]}t^{|A|}\det\Gam_A
 =\det(\id+t\Gam)=(1+t^2)^N,
 \qquad
 \sum_{|A|=2k}Y_A=\binom Nk.
 \label{eq:statewise-sector-mass}
\end{equation}
Consequently, for every pure Gaussian state, the half-weight $K=|A|/2$ obeys
\begin{equation}
 \Pr_\pi(K=k\mid\Gam)=2^{-N}\binom Nk,
 \qquad
 \Pr_{\widetilde\pi}(K=k\mid\Gam)=\frac{\binom Nk}{2^N-2},
 \quad 1\leq k\leq N-1,
 \label{eq:binomial-degree}
\end{equation}
where the first formula also includes $k=0,N$. In particular,
\begin{equation}
 \Pr_\pi(K=1\mid\Gam)=\frac{N}{2^N}.
 \label{eq:rare-mass}
\end{equation}
For $N\geq3$, the weight-two and weight-$(2N-2)$ sectors are distinct and together carry filtered sampling probability $p_\partial=2N/(2^N-2)$. They dominate the averaged filtered stabilizer purity for each fixed $q>2$, but the probability of seeing either of them in $M$ filtered draws is only
\begin{equation}
 p_{\mathrm{hit}}=1-(1-p_\partial)^M\leq Mp_\partial.
 \label{eq:sampling-hit-probability}
\end{equation}
A fixed nonzero hit probability therefore requires $M=\Theta(2^N/N)$. The sampling distribution places most draws at $K\simeq N/2$, whereas the high moment is carried by $K=1,N-1$.

The SRE density inferred from such samples can be computed directly. Draw a Majorana Haar state and then sample $M=M(N)\geq1$ strings from that same state. For a single filtered draw and any $\epsilon>0$, Markov's inequality gives
\begin{align}
 \Pr\!\left(Y>2^{-(1-\epsilon)N}\right)
 &\leq 2^{(1-\epsilon)N}\E\widetilde\zeta_2
 =O\!\left(N^{3/2}2^{-\epsilon N}\right),
 \label{eq:sampling-upper-tail}\\
 \Pr\!\left(Y<2^{-(1+\epsilon)N}\right)
 &\leq \frac{2^{2N-1}-2}{2^N-2}\,2^{-(1+\epsilon)N}
 =O\!\left(2^{-\epsilon N}\right).
 \label{eq:sampling-lower-tail}
\end{align}
The first line uses the exact second moment \cref{eq:maj-q2,eq:maj-filtered-exact}. The second follows by summing $Y_A/(D-2)$ over the at most $2^{2N-1}-2$ nontrivial even strings below the threshold. The $M$ draws share one random state, and the probability is taken jointly over the state and the samples. A union bound over the draws shows that, with probability $1-O(MN^{3/2}2^{-\epsilon N})$, every sampled value lies between the two thresholds. Substitution into \cref{eq:direct-sampling-estimator} then places $\widehat{\widetilde M}_q/N$ between $1-\epsilon$ and $1+\epsilon$. Thus, with sub-exponential number of samples, 
\begin{equation}
 \quad
 \frac{\widehat{\widetilde M}_q}{N}\xrightarrow{\mathbb P}1
 \quad(q>1),
 \qquad
 \frac{\widetilde M_q}{N}\xrightarrow{\mathbb P}\frac1{q-1}
 \quad(q>2).
 \quad
 \label{eq:sampling-false-density}
\end{equation}
Here $q$ is fixed and convergence is with respect to both the state and sampling randomness. The second limit is the exact typical-state density established in \cref{sec:quenched}. For $q>2$, the first limit is therefore a sampling artifact, although the stabilizer purity estimator is unbiased at every finite $N$.

The near-unit coefficient reported at $q=3$ in Ref.~\cite{Collura2026} is consistent with \cref{eq:sampling-false-density}. This limitation concerns the direct characteristic-sampling estimator; resolving low weights explicitly or using importance sampling can change its cost.

An independent deterministic bound makes this failure testable at finite size. Let $u_{ab}=\Gam_{ab}^2$ and $L=\binom{2N}{2}=N(2N-1)$. Since $\sum_{a<b}u_{ab}=N$, convexity implies, for $q\geq1$,
\begin{equation}
 \sum_{a<b}u_{ab}^{q}
 \geq L\left(\frac NL\right)^q
 =\frac{N}{(2N-1)^{q-1}}.
 \label{eq:power-mean-bound}
\end{equation}
Including the complementary sector gives, for every pure Gaussian state with $N\geq3$,
\begin{align}
 \widetilde M_q
 \leq \frac{\log[(2^N-2)/(2N)]}{q-1}+\log(2N-1),\qquad q>1.
 \label{eq:statewise-entropy-upper}
\end{align}
At $q=3$, this excludes any limiting SRE density greater than $1/2$ for any sequence of pure Gaussian states, independently of annealed or quenched averaging.

The unfiltered direct estimator has the same failure: with a subexponential budget, the protected pair of total sampling mass $2/D$ is also missed with probability tending to one, and the sampled SRE density tends to $1$. Its true value is $1/(q-1)$ for $q>2$. Adding the known protected contribution analytically gives
\begin{equation}
 \widehat\zeta_q^{\mathrm{corr}}=\frac2D+\frac{D-2}{D}\widehat{\widetilde\zeta}_q.
 \label{eq:unfiltered-corrected-estimator}
\end{equation}
For fixed $q>2$, the added term dominates the sampled bulk and restores the correct leading unfiltered density $1/(q-1)$. It does not recover the filtered SRE or the rare-string corrections.

\section{Number-conserving ensembles}
\label{sec:u1}

Particle-number conservation restricts the Gaussian orbit to the complex Grassmannian in \cref{eq:slater-ensemble}. Strings of equal Majorana weight are no longer equivalent under the ensemble symmetry. We instead average replicated states, obtaining an exact coefficient integral for integer moments and a separate expansion of the low-weight contributions.

\subsection{Replica reduction and the \texorpdfstring{$\mathrm{SU}(2q)$}{SU(2q)} coefficient integral}
\label{sec:u1-replica}

For a positive integer $q$, we introduce $2q$ identical copies of the state. The replicated Pauli operator
$Q_{2q}=(I^{\otimes2q}+X^{\otimes2q}+Y^{\otimes2q}+Z^{\otimes2q})^{\otimes N}$
collects the Pauli sum into the trace $\zeta_q=2^{-N}\Tr(\rho^{\otimes2q}Q_{2q})$~\cite{LeoneOlivieroZhouHamma2021,ZhuKuengGrasslGross2016,turkeshi2026lecturenotesreplicatensor,TurkeshiTirritoSierant2025,SierantTurkeshiTarabunga2026}. The Haar orbit of the replicated rank-$r$ Slater state spans one irreducible $\mathrm{U}(N)$ multiplet: the rectangle with $r$ rows of length $2q$. Its projector $\mathsf P^{(2q)}_{N,r}$ and dimension obey
\begin{equation}
 \E\rho_U^{\otimes2q}=\frac{\mathsf P^{(2q)}_{N,r}}{d_{N,r}^{(2q)}},
 \qquad
 d_{N,r}^{(2q)}=\prod_{j=0}^{2q-1}
 \frac{\Gamma(N+j+1)\Gamma(j+1)}{\Gamma(r+j+1)\Gamma(N-r+j+1)}.
 \label{eq:rectangle-dim}
\end{equation}
The first identity follows because Haar averaging is proportional to the identity on that multiplet, with its normalization fixed by unit trace. Consequently,
\begin{equation}
 \E\zeta_q^{\mathrm{U}(1)}
 =\frac{2^{-N}}{d_{N,r}^{(2q)}}\Tr[\mathsf P^{(2q)}_{N,r}Q_{2q}].
 \label{eq:u1-projector}
\end{equation}

The physical-mode and replica rotations act on different indices of the replicated fermionic Fock space. Skew Howe duality pairs the physical rectangle with a one-dimensional replica representation, $\det(s)^r$~\cite{Howe1989,Panova18skew}. This representation is invariant under $\mathrm{SU}(2q)$ and has total occupation $2qr$. Integrating the replica rotation and selecting this occupation therefore projects onto the multiplet in \cref{eq:u1-projector}. The result is
\begin{equation}
 \quad
 \E\zeta_q^{\mathrm{U}(1)}=\frac{1}{d_{N,r}^{(2q)}}[z^{2qr}]
 \int_{\mathrm{SU}(2q)}A_q(z,s)^N\dd s.
 \quad
 \label{eq:u1-all-q}
\end{equation}
Here $\dd s$ is normalized Haar measure and $[z^{2qr}]$ extracts the indicated coefficient. The contribution of one physical mode is
\begin{equation}
 \begin{split}
 A_q(z,s)&=\sum_{\ell=0,2,\ldots,2q}z^\ell e_\ell(s)+z^q\tau_q(s),\\
 \tau_q(s)&=\sum_{\substack{S\subset[2q]\\ |S|=q}}
 \epsilon_{S,S^c}\det s_{S^c,S}.
 \end{split}
 \label{eq:Aq}
\end{equation}
If $\lambda_1,\ldots,\lambda_{2q}$ are the eigenvalues of $s$, the elementary symmetric polynomial is
\begin{equation}
 e_\ell(s)=\sum_{i_1<\cdots<i_\ell}\lambda_{i_1}\cdots\lambda_{i_\ell}
 =\sum_{|S|=\ell}\det s_{S,S},
 \qquad e_0=1,\quad e_1=\Tr s,\quad e_{2q}=\det s=1.
 \label{eq:elementary-symmetric}
\end{equation}
Equivalently, $e_\ell$ is the trace of $s$ acting on the antisymmetric $\ell$-particle sector, denoted $\wedge^\ell\mathbb C^{2q}$. In $\tau_q$, the rows and columns of the minor are the complementary replica sets $S^c$ and $S$. The sign $\epsilon_{S,S^c}$ is the parity of the permutation that places the increasingly ordered elements of $S$ before those of $S^c$.

 On one physical mode, $I^{\otimes2q}+Z^{\otimes2q}$ retains the even replica occupations and contributes $2\sum_{\ell\,\mathrm{even}}z^\ell e_\ell(s)$. The $X$ and $Y$ insertions exchange an occupied set with its complement. Their trace survives only at occupation $q$ and contributes $2z^q\tau_q(s)$, including the fermionic shuffle sign. The local trace is thus $2A_q$; its $N$th power cancels the $2^{-N}$ in \cref{eq:u1-projector}.

Although the $\mathrm{SU}(2q)$ integral in \cref{eq:u1-all-q} has dimension $4q^2-1$, independent of $N$, coefficient extraction and cancellations still complicate the numerical evaluation of the ensemble-averaged stabilizer purity $\E\zeta_q^{\mathrm{U}(1)}$.
At $q=2$ a further reduction makes exact finite-size evaluation practical. For $q\geq3$, the complementary-minor term depends on eigenvectors as well as eigenvalues, so the integral does not reduce directly to an eigenangle integral.

\subsection{The second moment and its rate function}
\label{sec:u1-q2}

At $q=2$, the antisymmetric two-particle representation $\wedge^2\mathbb C^4$ admits a real form and realizes the double covering $\mathrm{SU}(4)\to\mathrm{SO}(6)$, with kernel $\{\pm I_4\}$. Consequently, normalized Haar averaging in this representation reduces to normalized Haar averaging over $\mathrm{SO}(6)$.
The elementary-symmetric and complementary-minor terms can then be combined, reducing \cref{eq:u1-all-q} to
\begin{equation}
 \begin{split}
 \E\zeta_2^{\mathrm{U}(1)}&=\frac{1}{d_{N,r}^{(4)}}
 \sum_{k=0}^{\min(r,N-r)}4^k\binom N{2k}\binom{N-2k}{r-k}\mu_{2k},\\
 \mu_{2k}&=\int_{\mathrm{SO}(6)}[\Tr(\Sigma_3O)]^{2k}\dd O,
 \qquad \Sigma_3=\diag(1,1,1,0,0,0).
 \end{split}
 \label{eq:u1-q2}
\end{equation}
The measure is normalized. The first values are $\mu_0=1$, $\mu_2=1/2$, $\mu_4=57/80$, $\mu_6=51/32$, and $\mu_8=597/128$. Their matrix-Bessel generating function gives finite rational evaluations of the sum. This expression agrees with the particle-number-conserving second moment of Ref.~\cite{Braccia2026}.

For $N\to\infty$ at fixed filling $\nu=r/N\in(0,1)$, we let $x_\nu>0$ andsolve $x_\nu(6+2x_\nu)/(1+6x_\nu+x_\nu^2)=2\nu$. With $H$ the binary entropy in bits, the exact exponential rate is
\begin{equation}
 I_2(\nu)=4H(\nu)-\log(1+6x_\nu+x_\nu^2)+2\nu\log x_\nu.
 \label{eq:I2}
\end{equation}
More precisely,
\begin{equation}
 \E\zeta_2^{\mathrm{U}(1)}\sim C(\nu)N^{3/2}2^{-NI_2(\nu)},
 \qquad M_{2,\mathrm{ann}}^{\mathrm{U}(1)}=NI_2(\nu)-\frac32\log N+O(1),
 \label{eq:u1-q2-rate}
\end{equation}
where $C(\nu)>0$. At half filling, $x_{1/2}=I_2(1/2)=1$. The Majorana and complex-Slater ensembles therefore have the same extensive and logarithmic terms at $q=2$.

\subsection{The diagonal sector and bounds on the purity decay rate}
\label{sec:u1-diagonal}

For $A\subseteq[N]$, we write $Z_A=\prod_{i\in A}Z_i$ and $\zeta_q^Z=2^{-N}\sum_A|\langle Z_A\rangle|^{2q}$. At fixed $|A|$ and filling, $\langle Z_A\rangle$ concentrates around $(1-2\nu)^{|A|}$. Retaining only $I$ and $Z$ insertions removes the complementary-minor term from the local replica trace. The same occupation projection as above gives, for positive integer $q$,
\begin{equation}
 \E\zeta_q^Z=\frac{1}{d_{N,r}^{(2q)}}[z^{2qr}]
 \int_{\mathrm{SU}(2q)}\left(\sum_{\ell=0}^q z^{2\ell}e_{2\ell}(s)\right)^N\dd s.
 \label{eq:diag-schur}
\end{equation}
This is the rectangular Schur coefficient: the group integral selects the determinant representation of degree $2qr$ in the symmetric-polynomial expansion. Its positivity also follows directly from the original sum over $Z$ strings.

We define $B_q(x)=\sum_{\ell=0}^q\binom{2q}{2\ell}x^\ell$ and choose $x_q>0$ so that $x_qB_q'(x_q)/B_q(x_q)=q\nu$. The diagonal contribution has rate
\begin{equation}
 J_q^Z(\nu):=-\lim_{N\to\infty}\frac1N\log\E\zeta_q^Z
 =2qH(\nu)-\log B_q(x_q)+q\nu\log x_q.
 \label{eq:JqZ}
\end{equation}
For the full moment, Jensen's inequality under the characteristic weights gives $\zeta_2^{q-1}\leq\zeta_q\leq\zeta_2$ for real $q\geq2$. Averaging over states gives $\E\zeta_q\geq(\E\zeta_2)^{q-1}$. Together with the protected strings and, for integer $q$, the diagonal contribution, these inequalities imply
\begin{equation}
 \begin{aligned}
 I_2(\nu)&\leq J_q(\nu)\leq\min\{(q-1)I_2(\nu),1\},
 &&q\in\mathbb R,\ q\geq2,\\
 I_2(\nu)&\leq J_q(\nu)\leq\min\{(q-1)I_2(\nu),1,J_q^Z(\nu)\},
 &&q\in\mathbb N,\ q\geq2,
 \end{aligned}
 \label{eq:rate-bracket}
\end{equation}
whenever $J_q(\nu)=-\lim N^{-1}\log\E\zeta_q^{\mathrm{U}(1)}$ exists. At half filling, the bounds coincide and prove $J_q(1/2)=1$ for every real $q\geq2$.

\subsection{The weight-two sector and half-filling asymptotics}
\label{sec:u1-weight2}

We now isolate the boundary contribution to the filtered moments. For $Q=\id-2G$, the weight-two entries are $Q_{ii}$ and, for each $i<j$, two copies each of the real and imaginary parts of $Q_{ij}=-2G_{ij}$, up to signs. The phase of $G_{ij}$ is uniform, and $G_{ii}\sim\operatorname{Beta}(r,N-r)$. Thus, for every real $q>0$,
\begin{equation}
 L_q(N,r):=\E\sum_{a<b}|\Gam_{ab}|^{2q}
 =N\E|1-2X|^{2q}+4\binom{2q}{q}\binom N2\E|G_{12}|^{2q},
 \quad X\sim\operatorname{Beta}(r,N-r).
 \label{eq:Lq-exact}
\end{equation}
The generalized binomial coefficient is $\binom{2q}{q}=\Gamma(2q+1)/\Gamma(q+1)^2$. For integer $q$, the diagonal beta moment is the terminating polynomial ${}_2F_1(-2q,r;N;2)$. The off-diagonal moment follows from another beta variable. The projector identity $G^2=G$ gives $\sum_{j\ne1}|G_{1j}|^2=X(1-X)$. Conditional on $X$, unitary invariance makes the remaining row a uniform direction in $\mathbb C^{N-1}$, so
\begin{equation}
 |G_{12}|^2\overset{\mathrm d}=X(1-X)T,
 \qquad T\sim\operatorname{Beta}(1,N-2),\quad T\ \text{independent of }X,
 \quad N>2.
 \label{eq:slater-radial-beta}
\end{equation}
At $N=2$, $T=1$. Averaging $X^q(1-X)^q$ and $T^q$ separately yields
\begin{equation}
 \E|G_{12}|^{2q}=\Gamma(q+1)
 \frac{\Gamma(r+q)\Gamma(N-r+q)}{\Gamma(r)\Gamma(N-r)}
 \frac{\Gamma(N-1)\Gamma(N)}{\Gamma(N+q-1)\Gamma(N+2q)},
 \label{eq:offdiag-selberg}
\end{equation}
valid for $N\geq2$, $1\leq r\leq N-1$, and real $q>0$.

At half filling, the beta and Gamma ratios give
\begin{align}
 L_q(N,N/2)&=a_qN^{2-q}\left[1+\frac{c_q}{N}+\frac{e_q}{N^2}+O(N^{-3})\right],
 \label{eq:Lq-half}\\
 a_q&=\frac{\Gamma(q+\tfrac12)}{\Gamma(\tfrac32)},\qquad
 c_q=2^{q-1}-1-\frac{q(q-1)}2,
 \label{eq:slater-a-c}\\
 e_q&=\frac{q(3q^3-2q^2+21q+2)}{24}-2^{q-1}q^2.
 \label{eq:app-eq}
\end{align}
Complementation gives an equal contribution at weight $2N-2$. The next correction comes from weight four: a generic minor has six asymptotically independent entries with variance $1/(2N)$, and its Pfaffian is a sum of three products. Its moment, multiplied by the $\binom{2N}{4}$ choices, gives
\begin{equation}
 \E\sum_{|A|=4}Y_A^q\sim a_qd_qN^{4-2q},
 \qquad d_q=\frac{a_q(q+\tfrac12)}{3}.
 \label{eq:slater-weight4}
\end{equation}
This term is needed because its relative size $N^{2-q}$ competes with the algebraic corrections to $L_q$: it gives the first correction for $2<q\leq3$, since $c_3=0$, and enters at order $N^{-2}$ for $q=4$.

In terms of $S_q^\circ=\sum_{P\notin\{I,Z^{\otimes N}\}}Y_P^q$, the boundary expansion takes the simple form
\begin{equation}
 \E S_q^\circ\simeq 2L_q(N,N/2)+2a_qd_qN^{4-2q}+\cdots .
 \label{eq:u1-filter-lower}
\end{equation}
Here $\simeq$ means that the fixed-weight contributions are summed under the assumption that growing-weight sectors remain subdominant. The displayed fixed-weight terms are those derived above; the assumption concerns only the sectors of growing weight. The stabilizer entropies then follow from the exact normalization identities
\begin{equation}
 M_{q,\mathrm{ann}}=\frac{N-\log(2+\E S_q^\circ)}{q-1},
 \qquad
 \widetilde M_{q,\mathrm{ann}}=\frac{\log(2^N-2)-\log\E S_q^\circ}{q-1}.
 \label{eq:slater-power-to-entropy}
\end{equation}
In particular, for fixed real $q>2$,
\begin{equation}
 \begin{aligned}
 M_{q,\mathrm{ann}}^{\mathrm{U}(1)}&\simeq\frac{N-1}{q-1}-\frac{a_q\loge}{q-1}N^{2-q}+\cdots,\\
 \widetilde M_{q,\mathrm{ann}}^{\mathrm{U}(1)}&\simeq
 \frac{N+(q-2)\log N-\log(2a_q)}{q-1}+\Delta_q^{\mathrm{U}(1)}(N),
 \end{aligned}
 \label{eq:u1-half-rate}
\end{equation}
where the first correction is
\begin{equation}
 \Delta_q^{\mathrm{U}(1)}(N)=\begin{cases}
 -d_q\loge\,N^{2-q}/(q-1),&2<q<3,\\
 -35\loge/(16N),&q=3,\\
 -c_q\loge/[(q-1)N],&q>3.
 \end{cases}
 \label{eq:u1-delta}
\end{equation}
Keeping the next terms at integer orders gives
\begin{align}
 M_{3,\mathrm{ann}}^{\mathrm{U}(1)}&\simeq\frac{N-1}{2}-\frac{15\loge}{8N}-\frac{75\loge}{16N^2}+O(N^{-3}),\\
 \widetilde M_{3,\mathrm{ann}}^{\mathrm{U}(1)}&\simeq\frac{N+\log N-\log(15/2)}2-\frac{35\loge}{16N}+O(N^{-2}),
 \label{eq:u1-q3-filter}\\
 \widetilde M_{4,\mathrm{ann}}^{\mathrm{U}(1)}&\simeq\frac{N+2\log N-\log(105/4)}3
 -\frac{\loge}{3N}+\frac{1085\loge}{48N^2}+O(N^{-3}).
 \label{eq:u1-q4-filter}
\end{align}
These subleading terms distinguish the complex-Slater and Majorana ensembles. The expansion is not uniform as $q\to2^+$; at the transition, \cref{eq:u1-q2-rate} applies. Independently of the boundary assumption, \cref{sec:quenched} proves the common extensive density.

Away from half filling, the mean diagonal entry is $1-2\nu$. A fixed subset containing $a$ doubly selected physical modes and $2t$ singly selected modes has $\Theta(N^{a+2t})$ realizations and typical Pfaffian size $O_{\mathbb P}(|1-2\nu|^aN^{-t/2})$. Its total annealed contribution therefore scales as
\begin{equation}
 \Theta\!\left(N^{a+t(2-q)}|1-2\nu|^{2qa}\right).
 \label{eq:kahler-hierarchy}
\end{equation}
This estimate applies as $N\to\infty$ with $a$, $t$, and $\nu\neq1/2$ held fixed. The integral in \cref{eq:diag-schur} gives the total contribution of Pauli strings containing only $I$ and $Z$, including those whose weight grows with $N$. The estimate above alone does not determine whether these diagonal strings dominate the ensemble-averaged stabilizer purity.

\section{The Pauli spectrum of \texorpdfstring{$\mathrm{SYK}_2$}{SYK2} ground states}
\label{sec:syk}

\subsection{Three quadratic ensembles}
\label{sec:syk-ensembles}

For the number-conserving quadratic Hamiltonian
\begin{equation}
 H_{\mathrm{SYK}_2}^{\mathrm{U}(1)}=\sum_{i,j=1}^N h_{ij}c_i^\dagger c_j,
 \qquad h=h^\dagger,
 \label{eq:syk-number-conserving}
\end{equation}
the ground state at fixed particle number $r$ fills the $r$ lowest one-particle orbitals. If $h$ belongs to the GUE, its eigenvectors generate the complex Slater ensemble; for GOE hopping, they generate the real-Slater ensemble. The eigenvalues determine the occupied orbitals but do not affect their invariant distribution~\cite{LydzbaRigolVidmar2020}. Throughout the numerical comparison below, $N$ counts complex fermionic modes and $r=N/2$.

For the pairing-allowed Hamiltonian,
\begin{equation}
 H_{\mathrm{SYK}_2}^{\mathrm{Maj}}=\frac{i}{4}\sum_{a,b=1}^{2N}A_{ab}\gamma_a\gamma_b,
 \qquad A=-A^{\mathsf T}\in\mathbb R^{2N\times2N},
 \label{eq:syk-majorana-orbit}
\end{equation}
orthogonally invariant Gaussian couplings give a uniformly distributed orthogonal eigenframe. Conditional on fermion parity, the ground state is a Majorana Haar state; both parity components have the same squared Pauli spectrum and SREs. Thus the Majorana results apply to this quadratic ensemble~\cite{Collura2026}.

\subsection{Real Slater determinants}
\label{sec:syk-real}

For a real projector $G=G^{\mathsf T}$, let $Q=\id-2G$ and order the Majoranas by species, $\eta_i=\gamma_{2i-1}$ followed by $\chi_i=\gamma_{2i}$. Then
\begin{equation}
 \Gam=\begin{pmatrix}0&Q\\-Q&0\end{pmatrix},
 \qquad QQ^{\mathsf T}=\id.
 \label{eq:real-covariance}
\end{equation}
A string with $\eta$ indices $I$ and $\chi$ indices $J$ vanishes unless $|I|=|J|$; its magnitude is then $|\det Q_{I,J}|$. At half filling these balanced minors are generically nonzero, so the size of the support, including identity and parity, is
\begin{equation}
 \sum_{k=0}^N\binom Nk^2=\binom{2N}{N}.
 \label{eq:real-support}
\end{equation}
Away from half filling, rank constraints impose additional zeros when $|I|-|I\cap J|>\min(r,N-r)$. 
To verify normalization at arbitrary filling, we apply the Cauchy--Binet identity to $QQ^{\mathsf T}=\id$:
\begin{equation}
 \sum_{|J|=k}\det Q_{I,J}^2=\det[(QQ^{\mathsf T})_{I,I}]=1,
 \qquad
 \sum_{k=0}^N\sum_{|I|=|J|=k}\det Q_{I,J}^2=2^N.
 \label{eq:real-parseval}
\end{equation}

To calculate the contribution from strings of Majorana weight two, we need the distributions of the diagonal and off-diagonal entries of $Q$. A diagonal entry has the same distribution as $1-2X$, where $X\sim\operatorname{Beta}(r/2,(N-r)/2)$. A squared off-diagonal entry has the same distribution as $4X(1-X)B$, where $B\sim\operatorname{Beta}(1/2,(N-2)/2)$ is independent of $X$ for $N>2$. At $N=2$, we set $B=1$. Averaging $|Q_{ij}|^{2q}$ and summing over the matrix entries gives, for positive integer $q$,
\begin{equation}
 L_q^{\mathbb R}(N,r)
 =N\,{}_2F_1\!\left(-2q,\frac r2;\frac N2;2\right)
 +N(N-1)4^q\frac{\rising{r/2}{q}\rising{(N-r)/2}{q}\rising{1/2}{q}}
 {\rising{N/2}{2q}\rising{(N-1)/2}{q}}.
 \label{eq:real-L}
\end{equation}
For real noninteger $q>0$, the first term is instead $N\E|1-2X|^{2q}$; the second term is unchanged. At half filling,
\begin{equation}
 L_q^{\mathbb R}(N,N/2)=(2q-1)!!\,N^{2-q}
 \left[1+\frac{2^q-q^2-1}{N}+O(N^{-2})\right]
 \label{eq:real-L-half}
\end{equation}
for integer $q>0$. The leading coefficient is $2^{q-1}a_q$, reflecting the broader distribution of the real matrix elements.

For fixed weight $2k$, the dominant disjoint balanced minors approach $k\times k$ real Gaussian determinants after multiplying their entries by $\sqrt N$. Their determinant moments and multiplicity give
\begin{equation}
 T_{k,q}^{\mathbb R}:=\E\sum_{|A|=2k}Y_A^q
 \sim\frac{2^{kq}}{(k!)^2}\prod_{j=1}^k\frac{\Gamma(q+j/2)}{\Gamma(j/2)}\,N^{k(2-q)}.
 \label{eq:real-fixed-weight}
\end{equation}
As in \cref{sec:u1-weight2}, applying the fixed-weight series to the full moment assumes boundary dominance. For integer $q\geq4$, this gives
\begin{equation}
 \widetilde M_{q,\mathrm{ann}}^{\mathbb R}\simeq
 \frac{N+(q-2)\log N-\log[2(2q-1)!!]+(q^2+1-2^q)\loge/N}{q-1}.
 \label{eq:real-filtered}
\end{equation}
At $q=3$, $T_{2,3}^{\mathbb R}\sim180N^{-2}$ contributes at the same order as the correction to $L_3^{\mathbb R}=15N^{-1}(1-2/N+O(N^{-2}))$. Including both complementary boundaries gives
\begin{equation}
 \widetilde M_{3,\mathrm{ann}}^{\mathbb R}\simeq
 \frac12\left[N+\log N-\log30-\frac{10\loge}{N}+O(N^{-2})\right].
 \label{eq:real-q3}
\end{equation}

\subsection{Ensemble discrimination at subleading order}
\label{sec:syk-discrimination}

At $q=3$, the common leading expression for the Majorana and complex-Slater filtered SREs is $\tfrac12[N+\log N-\log(15/2)]$. Their coefficients of $\loge/N$ are $-19/16$ and $-35/16$, respectively. For real Slater states, the leading filtered power sum is four times larger: the SRE is lower by one bit at constant order, with coefficient $-5$ for $\loge/N$; see \cref{eq:maj-q3-filtered,eq:u1-q3-filter,eq:real-q3}. The Majorana expansion is asymptotically exact, while the Slater subleading terms rely on the boundary-dominance assumption. The common extensive coefficient is proved for Majorana and half-filled complex Slater states in \cref{sec:quenched}.

\subsection{Bulk versus rare strings}
\label{sec:syk-support}

We now use the Majorana product law in \cref{eq:beta-product} and the complex Slater distribution in \cref{eq:slater-radial-beta} to motivate an approximate description of the typical Pauli expectation values in half-filled complex $\mathrm{SYK}_2$ ground states.
To see this, we first separate the factor $B_0$ from the Majorana product law \cref{eq:beta-product}. For $m=\min(k,N-k)\geq1$, the beta-to-Gaussian limit $2(N-m)B_0\Rightarrow Z^2$, with $Z\sim\mathcal N(0,1)$, gives the signed amplitude in the form
\begin{equation}
 x_A\simeq\sqrt{V_{N,k}}\,Z,
 \qquad V_{N,k}=\frac{1}{2(N-m)}\prod_{j=1}^{m-1}B_j,
 \qquad Z\ \text{independent of }V_{N,k}.
 \label{eq:beta-gaussian-mixture}
\end{equation}
Thus the beta product separates into a Gaussian amplitude and a fluctuating variance.

For complex Slater states, the radial beta variable in \cref{eq:slater-radial-beta} gives a simple variance fluctuation: $(N-1)T\Rightarrow W$, with $W\sim\operatorname{Exp}(1)$. We approximate the accumulated variance of a bulk string by one such effective complex intensity, preserving the scale-mixture form of \cref{eq:beta-gaussian-mixture}. This replacement is the approximation step of our bulk description:
\begin{equation}
 x_P\simeq\sqrt{V_N}\,Z,
 \qquad V_N=\sigma_N^2 W,
 \qquad p_{V_N}(v)=\frac{e^{-v/\sigma_N^2}}{\sigma_N^2},\quad v>0,
 \label{eq:slater-effective-variance}
\end{equation}
where $Z$ and $W$ are independent and $\sigma_N^2$ is the mean squared amplitude. Averaging the conditional Gaussian over this variance gives
\begin{equation}
 \begin{split}
 f_N(x)&=\int_0^\infty\frac{\exp[-x^2/(2v)]}{\sqrt{2\pi v}}
               \frac{e^{-v/\sigma_N^2}}{\sigma_N^2}\dd v
       =\frac{e^{-|x|/b_N}}{2b_N},\\
 b_N^2&=\frac{\sigma_N^2}{2}.
 \end{split}
 \label{eq:slater-exponential-ansatz}
\end{equation}
The exponential profile therefore follows from retaining one effective complex radial fluctuation in the beta-based description. Its logarithm is linear in $-|x|$, which is the triangular semilogarithmic profile observed in Ref.~\cite{BeraSchiro2025}.

\begin{figure}[t]
 \centering
 \includegraphics[width=0.7\linewidth]{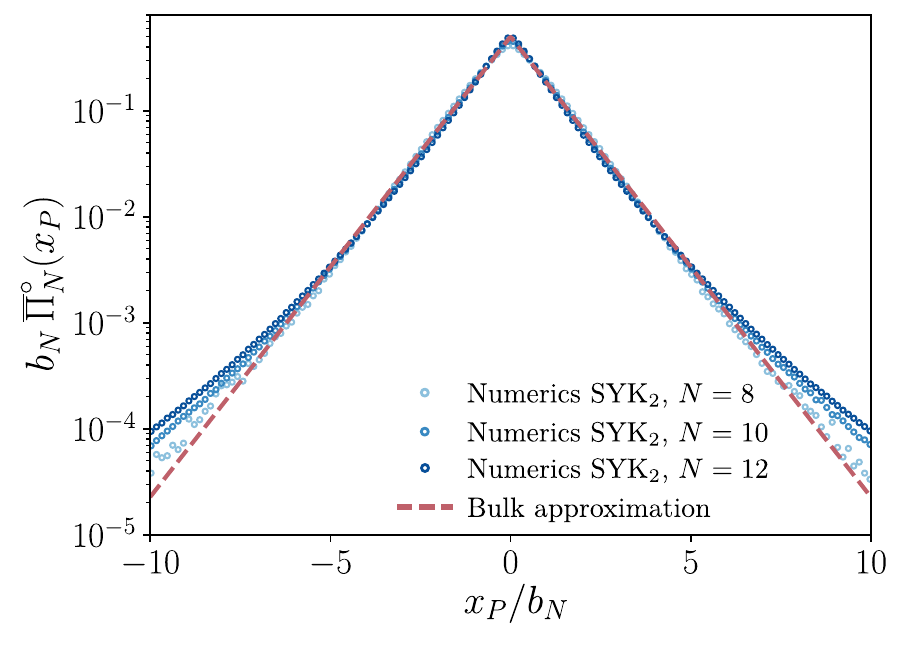}
 \caption{Signed Pauli spectrum of half-filled GUE $\mathrm{SYK}_2$ ground states, conditioned on nontrivial even strings. Symbols: all such Pauli expectations at $N=8,10,12$, evaluated from the occupied-orbital covariance matrices using Wick's theorem. Dashed curve: the analytical bulk approximation \cref{eq:slater-exponential-ansatz}, obtained by averaging Gaussian amplitudes over an effective exponential variance, with $b_N$ fixed by Parseval's identity. Both axes are rescaled by $b_N$.}
 \label{fig:u1-laplace}
\end{figure}

The remaining scale is fixed by the statewise normalization. For the complex half-filled ensemble, we retain the $n_N=2^{2N-1}-2$ nontrivial even strings with uniform weight and write
\begin{equation}
 \Pi_N^\circ(x)=\frac1{n_N}\sum_{\substack{P\ne I,Z^{\otimes N}\\P\ \mathrm{even}}}\delta(x-x_P),
 \qquad \sigma_N^2=\int x^2\Pi_N^\circ(x)\dd x=\frac{D-2}{n_N}.
 \label{eq:slater-conditional-spectrum}
\end{equation}
Here ``even'' refers to Majorana weight. Parseval's identity thus gives $b_N^2=(D-2)/(2n_N)$, with no adjustable width. \Cref{fig:u1-laplace} compares this prediction with every nontrivial even Pauli expectation in the GUE $\mathrm{SYK}_2$ ground-state samples at $N=8,10,12$.

The corresponding filtered moments are
\begin{equation}
 \widetilde\zeta_q^{\mathrm{exp}}=\frac{\Gamma(2q+1)}{2^q}
 \left(\frac{D-2}{n_N}\right)^{q-1},\qquad q>0.
 \label{eq:laplace-moments}
\end{equation}
The low-weight analysis predicts enhanced high moments relative to this bulk approximation. At $q=2$, the full moment contains the $N^{3/2}$ enhancement of \cref{eq:u1-q2-rate}; for fixed $q>2$, the boundary expansion exceeds \cref{eq:laplace-moments} by a factor $\Theta(N^{2-q}2^{(q-2)N})$.

\section{Annealed versus quenched nonstabilizerness}
\label{sec:quenched}

The annealed SRE is obtained by averaging the stabilizer purity over states before taking its logarithm. This average could, in principle, be strongly influenced by atypical states with unusually large purity. This possibility is distinct from a small subset of Pauli strings dominating the purity within a given state. To determine whether the annealed result describes typical states, we compare it with the quenched SRE, obtained by averaging the entropy of individual states, and examine fluctuations across the ensemble. For the Majorana Haar and half-filled complex Slater ensembles, we show that both averages have leading term $N/(q-1)$ and that individual SRE densities approach $1/(q-1)$ with probability tending to one, for every fixed real $q\geq2$.

We recall that $S_q^\circ=(2^N-2)\widetilde\zeta_q$. The normalization $\sum_{a<b}\Gam_{ab}^2=N$, convexity, and the complementary weight-two sector give the statewise bound already used in \cref{sec:freezing-sampling}:
\begin{equation}
 S_q^\circ\geq2\sum_{a<b}|\Gam_{ab}|^{2q}
 \geq2N(2N-1)^{1-q},\qquad q\geq1,\quad N\geq3.
 \label{eq:Sq-lower-statewise}
\end{equation}
Since every squared expectation lies in $[0,1]$, we also have $S_q^\circ\leq S_2^\circ$ for $q\geq2$.

We consider either the Majorana Haar ensemble or the complex Slater ensemble at half filling, with $N$ even in the latter case. 
For every fixed real $q\geq2$, the quenched SREs satisfy the following asymptotic relations, which we prove below:
\begin{equation}
 M_{q,\mathrm{que}}=\frac{N}{q-1}+O(\log N),
 \qquad
 \widetilde M_{q,\mathrm{que}}=\frac{N}{q-1}+O(\log N),
 \label{eq:quenched-half-density}
\end{equation}
so the quenched SREs have the same extensive coefficients as their annealed counterparts. Moreover, for every $\epsilon>0$, there exists a constant $\Lambda_q>0$ such that, for all sufficiently large $N$,
\begin{equation}
 \Pr\biggl(\Bigl|\frac{\widetilde M_q}{N}-\frac{1}{q-1}\Bigr|>\epsilon\biggr)
 \leq \Lambda_q\,N^{3/2}\,2^{-(q-1)\epsilon N/2}.
 \label{eq:quenched-concentration}
\end{equation}

The quenched SRE formulas in \cref{eq:quenched-half-density} and the concentration bound in \cref{eq:quenched-concentration} follow from the bounds valid for every state given above and the exact results for the ensemble-averaged stabilizer purity at $q=2$.
Indeed, for real $q\geq2$, the inequality
$S_q^\circ\leq S_2^\circ$ holds statewise, while \cref{eq:maj-q2,eq:u1-q2,eq:u1-q2-rate} at half filling give
$\E S_2^\circ=\Theta(N^{3/2})$. Jensen's inequality therefore yields
\begin{equation}
 \E\log S_q^\circ
 \leq\log\E S_q^\circ
 \leq\log\E S_2^\circ
 =O(\log N).
\end{equation}
Conversely, \cref{eq:Sq-lower-statewise} gives
$\log S_q^\circ\geq\log[2N(2N-1)^{1-q}]=-O(\log N)$ statewise. Hence
$\E\log S_q^\circ=O(\log N)$ in the two-sided sense. Since
\begin{equation}
 \widetilde M_q
 =\frac{\log(D-2)-\log S_q^\circ}{q-1},
\end{equation}
the filtered statement in \cref{eq:quenched-half-density} follows. Replacing $S_q^\circ$ by $2+S_q^\circ$ proves the unfiltered statement. The same polynomial upper and lower bounds imply
$\log\E S_q^\circ=O(\log N)$ and therefore establish the corresponding annealed densities directly, without relying on the unproved dominance assumption in the complex-Slater boundary expansion of \cref{eq:u1-filter-lower}.

For completeness, we set
$\delta_N=N-\log(D-2)=-\log(1-2^{1-N})=O(2^{-N})$. Then
\begin{equation}
 \frac{\widetilde M_q}{N}-\frac{1}{q-1}
 =-\frac{\log S_q^\circ+\delta_N}{(q-1)N}.
\end{equation}
For all sufficiently large $N$, the deviation event in \cref{eq:quenched-concentration} is contained in the union of
\begin{equation}
 S_q^\circ>2^{(q-1)\epsilon N/2}
 \qquad\text{and}\qquad
 S_q^\circ<2^{-(q-1)\epsilon N/2}.
\end{equation}
The second event is excluded by the polynomial statewise lower bound \cref{eq:Sq-lower-statewise}. For the first, Markov's inequality and $S_q^\circ\leq S_2^\circ$ give
\begin{equation}
 \Pr\!\left(S_q^\circ>2^{(q-1)\epsilon N/2}\right)
 \leq2^{-(q-1)\epsilon N/2}\,\E S_2^\circ
 \leq\Lambda_qN^{3/2}2^{-(q-1)\epsilon N/2},
\end{equation}
which proves \cref{eq:quenched-concentration}.

Thus, in the thermodynamic scaling limit, the magic density self-averages and the annealed prediction is exact for typical states in these two ensembles. This statement holds for each fixed real $q\geq2$; it does not identify the subleading finite-size corrections of the quenched SREs.

\section{Magic under partial trace}
\label{sec:partial-trace}

We now investigate how discarding modes affects the magic of states drawn from the Majorana Haar ensemble. The initial pure Gaussian state $|\psi_G\rangle$ has covariance matrix $\Gam=OJO^{\mathsf T}$, with $O$ Haar distributed on $\mathrm{SO}(2N)$. Tracing out the rightmost $N-n$ modes gives the reduced state
\begin{equation}
\rho_n
=
\Tr_{{n+1,\ldots,N}}
|\psi_G\rangle\!\langle\psi_G|,
\qquad
n=cN,
\qquad
\kappa=1-c.
\end{equation}
 This choice ensures that the Majorana strings supported on the retained subsystem map to local Pauli strings under the Jordan--Wigner transformation.
For an $n$-qubit mixed state, define the identity-filtered stabilizer norm~\cite{HowardCampbell2017}
\begin{equation}
\widetilde{\mathcal D}(\rho)
=
\frac{1}{2^n-1}
\sum_{\substack{P\in\Pcal_n\\P\ne\id}}
\left|\Tr(\rho P)\right|.
\label{eq:pt-Dnorm-def}
\end{equation}
The condition $\widetilde{\mathcal D}(\rho)>1$ certifies that $\rho$ lies outside the stabilizer polytope, and it provides a lower bound on a mixed-state magic monotone called the Robustness of Magic~\cite{HowardCampbell2017}. 
One may similarly introduce filtered R\'enyi magic witnesses for $q\geq1/2$~\cite{TurkeshiDymarskySierant2025,HaugTarabunga2026}. Whenever any of these witnesses certifies magic, the stabilizer-norm condition $\widetilde{\mathcal D}(\rho)>1$ also holds. We therefore focus on $\widetilde{\mathcal D}$.
Notice that the appropriate filtration is now identity-only: after tracing out modes, the parity operator of the reduced density matrix is no longer pinned.
\subsection{The reduced-state stabilizer norm}
\label{sec:pt-moments}
For every Majorana string $A\subseteq[2n]$ supported on the retained modes,
\begin{equation}
\Tr(\rho_n\gamma_A)
=
\langle\psi_G|\gamma_A|\psi_G\rangle.
\end{equation}
Odd strings vanish, while for $|A|=2r$ the even correlator is a principal Pfaffian whose first absolute moment is $R_{N,r}(1/2)$ from \cref{eq:R-general}. Therefore,
\begin{equation}
\E \widetilde{\mathcal D}(\rho_n)
=
\frac{1}{2^n-1}
\sum_{r=1}^{n}
\binom{2n}{2r}R_{N,r}(1/2).
\label{eq:pt-Dnorm-exact}
\end{equation}
Thus the average filtered stabilizer norm follows directly from the sector law already derived for the pure state. To obtain its leading exponential behavior, set $r=\alpha N$. Using \cref{eq:R-general} together with Stirling's formula gives
\begin{equation}
\begin{aligned}
\frac1N\log\binom{2n}{2r}
&=
2cH\!\left(\frac{\alpha}{c}\right)+o(1),\\
\frac1N\log R_{N,\alpha N}(1/2)
&=
-\frac12H(\alpha)+o(1).
\end{aligned}
\end{equation}
The complement symmetry of the Pfaffian moments extends the second expression to the full range $0\leq\alpha\leq1$. Hence,
\begin{equation}
\frac1N
\log
\sum_{r=1}^{n}
\binom{2n}{2r}R_{N,r}(1/2)
=
F(c)+o(1),
\label{eq:pt-logsum}
\end{equation}
where the rate is the sector exponent evaluated at its optimal Majorana weight,
\begin{equation}
F(c)
=
2cH\!\left(\frac{\alpha_\star}{c}\right)
-
\frac12H(\alpha_\star).
\label{eq:pt-Fdef}
\end{equation}
Here $\alpha_\star=\alpha_\star(c)\in(0,c)$ is the unique interior maximizer of the sector exponent $2cH(\alpha/c)-\tfrac12H(\alpha)$ over $0\leq\alpha\leq c$, equivalently the unique solution of the stationarity condition
\begin{equation}
4\log\frac{c-\alpha_\star}{\alpha_\star}
=
\log\frac{1-\alpha_\star}{\alpha_\star},
\qquad\text{i.e.}\qquad
(c-\alpha_\star)^4=\alpha_\star^{3}(1-\alpha_\star).
\label{eq:pt-saddle}
\end{equation}
 The annealed stabilizer-norm density is therefore
\begin{equation}
\lim_{N\to\infty}\frac{2}{N}
\log\E\widetilde{\mathcal D}(\rho_n)
=
2\bigl[F(c)-c\bigr].
\label{eq:pt-annealed-density}
\end{equation}
As in \cref{sec:quenched}, the annealed exponent is also the typical exponent. Define the unnormalized filtered norm
\begin{equation}
S_n
=
\sum_{\substack{P\in\Pcal_n\\P\ne\id}}
\left|\Tr(\rho_nP)\right|
=
(2^n-1)\widetilde{\mathcal D}(\rho_n).
\end{equation}
For every fixed $0<c<1$,
\begin{equation}
\frac1N\log S_n
\xrightarrow{\mathbb P}
F(c),
\qquad
\frac1N\E\log S_n
\longrightarrow
F(c).
\label{eq:pt-Sn-limit}
\end{equation}
More precisely, for every $\epsilon>0$ there is an $\eta_{\epsilon,c}>0$ such that
\begin{equation}
\Pr\left[
\left|
\frac1N\log S_n-F(c)
\right|>\epsilon
\right]
\leq
2^{-\eta_{\epsilon,c}N+o(N)}.
\label{eq:pt-Sn-concentration}
\end{equation}
The proof combines the first moment with the negative moments of the sector law. By \cref{eq:pt-logsum}, $\E S_n=\sum_r\binom{2n}{2r}R_{N,r}(1/2)=2^{NF(c)+o(N)}$, so Markov's inequality gives the upper tail $\Pr[N^{-1}\log S_n>F(c)+\epsilon]\leq2^{-\epsilon N+o(N)}$, while Jensen's inequality gives $\E\log S_n\leq\log\E S_n=NF(c)+o(N)$.

For a matching lower bound on the mean, fix a sector sequence $r=r_N$ with $r/N\to\alpha$, set $M_r=\binom{2n}{2r}$, and restrict $S_n$ to this sector. The arithmetic--geometric mean inequality then gives
\begin{equation}
\log S_n
\geq
\log M_r
+
\frac1{M_r}
\sum_{\substack{A\subseteq[2n]\\|A|=2r}}
\log|x_A|.
\label{eq:pt-amgm}
\end{equation}
The summands are identically distributed by rotational invariance, and \cref{eq:beta-product,eq:log-cumulants} give $\E\log|x_A|=-\tfrac N2H(\alpha)+O(\log N)$. Taking expectations in \cref{eq:pt-amgm} and choosing $\alpha=\alpha_\star(c)$ yields $\E\log S_n\geq NF(c)-O(\log N)$, which with the Jensen bound proves the convergence of $N^{-1}\E\log S_n$ in \cref{eq:pt-Sn-limit}.

The lower tail requires no independence between strings. Continuing the sector law to $-\tfrac12<q<0$ gives the negative moments
\begin{equation}
\E Y_A^{-s}=R_{N,r}(-s)=2^{sNH(\alpha)+o(N)},\qquad 0<s<\tfrac12,
\end{equation}
with $s<\tfrac12$ fixed by the smallest beta shape parameter, so Markov gives $\Pr[Y_A<2^{-N[H(\alpha)+\delta]}]\leq2^{-s\delta N+o(N)}$ for every $\delta>0$. Let $Z_r$ be the fraction of sector strings violating this bound; linearity of expectation and Markov give $\Pr[Z_r>\tfrac12]\leq2\E Z_r\leq2^{-s\delta N+o(N)}$. On the complementary event at least half the strings obey $|x_A|\geq2^{-N[H(\alpha)+\delta]/2}$, so $S_n\geq\tfrac{M_r}2\,2^{-N[H(\alpha)+\delta]/2}$; with $\alpha=\alpha_\star(c)$ this gives $N^{-1}\log S_n\geq F(c)-\tfrac\delta2+o(1)$ off an event of probability $\leq2^{-s\delta N+o(N)}$. Choosing $\delta<2\epsilon$ and combining with the upper tail proves \cref{eq:pt-Sn-concentration}.
 
The logarithmic stabilizer-norm density vanishes when $F(c)=c$, whose nontrivial solution is
\begin{equation}
c_\star
=
0.2387413856\ldots,
\qquad
\kappa_\star^{\rm G}
=
1-c_\star
=
0.7612586144\ldots .
\label{eq:pt-cstar}
\end{equation}
For $\kappa<\kappa_\star^{\rm G}$, the reduced Gaussian state is therefore certified to be magical with probability approaching one. For $\kappa>\kappa_\star^{\rm G}$, the filtered stabilizer norm falls below the witness threshold with high probability; this only means that it ceases to certify magic, not that the state is necessarily a stabilizer mixture.
\subsection{Comparison with Haar-random states}
\label{sec:pt-haar}
It is instructive to compare the Gaussian certification threshold $\kappa_\star^{\rm G}$ with the magic transition of a Haar-random state under partial trace. Recent results~\cite{LiuLiu2026} on the geometry of the stabilizer polytope show that a Haar-induced state on dimension $d$, obtained with an environment of dimension $K$, undergoes a magic transition at
\begin{equation}
\Omega\!\left(\frac{d^2}{\log^2d}\right)
\leq
K_\star
\leq
O(d^2).
\label{eq:pt-Kstar}
\end{equation}
On the exponential scale relevant here,
\begin{equation}
K=2^{\kappa N},
\qquad
d=2^{(1-\kappa)N},
\end{equation}
so $K_\star=d^{2+o(1)}$ corresponds to
\begin{equation}
\kappa_\star^{\rm Haar}
=
\frac23.
\end{equation}
Away from the transition window, the Haar-induced state is magical with high probability for $\kappa<2/3$, and belongs to the stabilizer polytope with high probability for $\kappa>2/3$.
Combining this result with \cref{eq:pt-cstar} yields the interval
\begin{equation}
\frac23
<
\kappa
<
0.7612586144\ldots,
\label{eq:pt-interval}
\end{equation}
in which the reduced Haar-random state is typically a stabilizer mixture, while the reduced random Gaussian state is still certified to be magical with high probability. This ordering runs counter to the usual intuition. Haar-random states are the paradigm of maximal quantum complexity, whereas fermionic Gaussian states are classically simulable and confined to an exponentially small submanifold. One would therefore naively expect the magic of Gaussian states to be the more fragile of the two. Instead, it is the structurally simpler ensemble whose magic survives partial trace over a strictly larger range of traced fractions, so that by this measure the ostensibly more complex Haar-random states are the less robust.

\section{Discussion and conclusions}
\label{sec:discussion}
\label{sec:conclusions}
Stabilizer R\'enyi entropies describe magic through moments of the squared Pauli expectation values. The index $q$ determines how strongly the largest values contribute relative to the rest of the spectrum.
With increasing mode number $N$, the exact Majorana spectrum separates exponentially many central strings from polynomially many low-weight strings whose expectation values are only algebraically small. Increasing $q$ transfers the moment to the boundaries at $q_c=2$, producing freezing that survives removal of contributions of identity and parity operators. We obtain the exact finite-size Majorana moments, a coefficient integral for every positive integer complex-Slater moment, and exact real-Slater weight-two moments. The symmetry class of the quadratic couplings determines which of these ensembles the $\mathrm{SYK}_2$ ground states form.

For $q>2$, Majorana and half-filled complex Slater states have filtered magic density $1/(q-1)$ in both annealed and quenched averages. In the Majorana ensemble, perfect characteristic sampling with a subexponential budget yields an apparent unit density, because it misses the rare strings carrying the moment. The statewise bound can be checked at finite size, and the sector formulas point to estimators that resolve the low-weight sectors explicitly. The beta-based variance-mixture approximation reproduces the triangular $\mathrm{SYK}_2$ bulk through $N=12$, while the low-weight sectors explain the enhancement of its high moments.
A related separation between the spectral bulk and enhanced contributions from low-weight operators has been identified in chaotic equilibrium states~\cite{Sarma26uni}, where thermal expectation values govern the high-order stabilizer moments. For random fermionic Gaussian states, the hierarchy is organized by Majorana weight and follows from Wick's theorem. Both settings illustrate how a small fraction of the Pauli spectrum can control high-order stabilizer entropies even after filtering.

These pure-state results can also be turned into a statement about mixed-state magic on traced out modes. The identity-filtered stabilizer norm of the reduced Gaussian state is governed by the same sector law, and both its annealed and typical densities stay positive---so the reduced state is certified to be magical---for every traced fraction below $\kappa_\star^{\rm G}\approx0.761$. This threshold exceeds the $\kappa=2/3$ at which a Haar-random state enters the stabilizer polytope~\cite{LiuLiu2026}, so random Gaussian states retain certifiable magic under partial trace over a strictly larger range than the generic states usually regarded as maximally complex. Whether $\kappa_\star^{\rm G}$ marks the true onset of stabilizer-mixture behavior, rather than only the failure of the witness, remains open, as does the corresponding threshold for the number-conserving ensembles.

The main unresolved analytical steps concern number-conserving states: controlling the growing-weight sectors behind the subleading boundary expansion, determining the full rate away from half filling, and evaluating the real-Slater second-moment asymptotics. For the complex coefficient integral, the complementary-minor term introduces eigenvector dependence and cancellations. A positive spectral representation analogous to the Majorana beta-product mixture would make the sector structure of the moments explicit and simplify their numerical evaluation.

Beyond the extensive SRE, mixed purity moments could determine quenched subleading terms and state-to-state fluctuations. In monitored Gaussian dynamics~\cite{TirritoEtAl2025,LumiaTirritoFazioCollura2024,wang2025magictransitionmonitoredfree} and doped matchgate circuits~\cite{PaviglianitiEtAl2026,BallarTriguerosEtAl2026,tyw6-pjp6,aditya2026equivalencequantumresourcesergodic,79vj-nx6r}, the rare-string contribution could be followed in time. Comparing it with fermionic non-Gaussianity measures~\cite{SierantStornatiTurkeshi2026,Lyu24ferm,Lyu25displ,HaugTurkeshiSierant2026,TarabungaEtAl2026,Tarabunga2026,Turkeshi2026,Ares26asym,ares2026nongaussianityrandomquantumstates} would separate motion within the Gaussian orbit from departures from Gaussianity.

\paragraph{Acknowledgments.}
We thank Ra{\'u}l Morral-Yepes and Marc Langer for useful discussions.
X.T. acknowledges support from DFG Emmy Noether Programme proposal
``Digital Quantum Matter Out-of-Equilibrium'' No. 560726973, DFG under
Germany's Excellence Strategy -- Cluster of Excellence Matter and Light for
Quantum Computing (ML4Q) EXC 2004/2 -- 390534769, and DFG Collaborative
Research Center (CRC) 183 Project No. 277101999 -- project B01.
P.S. acknowledges a fellowship within the ``Generaci\'on D'' initiative,
Red.es, Ministerio para la Transformaci\'on Digital y de la Funci\'on
P\'ublica, for talent attraction (C005/24-ED CV1), funded by the European
Union NextGenerationEU funds, through PRTR. P.S.T. acknowledges funding from the European Union (ERC, DynaQuant, No. 101169765).
\paragraph{Data availability.}
Numerical implementation and data will be shared publicly at publication.

\appendix

\section{Jacobi ensembles and the Selberg sector law}
\label{app:selberg}

In this appendix, we derive the sector law \cref{eq:R-general} and the moment formula \cref{eq:maj-all-q}. Fix $A\subseteq[2N]$ with $|A|=2k$, and first suppose $1\leq k\leq N/2$. The principal submatrix $\Gam_A$ of $\Gam=OJO^{\mathsf T}$ is the compression of a Haar-rotated complex structure to $2k$ coordinate directions. Its $k$ positive singular values may be written as $\{\sqrt{\xi_j}\}_{j=1}^{k}$, with $\xi_j\in[0,1]$. They form the Jacobi ensemble familiar from fermionic Page-curve calculations~\cite{Bianchi2021}, with joint density given by~\cref{eq:app-jacobi}.
In Selberg notation, the parameters are
$(\alpha,\beta,\gamma)=(\tfrac12,\,N-2k+1,\,1)$. Since
$Y_A=\det\Gam_A=\prod_j\xi_j$, inserting $Y_A^q$ shifts $\alpha$ to $\alpha+q$. The moment is therefore the ratio of Selberg integrals~\cite{Selberg1944,ForresterWarnaar2008,Forrester2010}
\begin{equation}
 \E[Y_A^q]
 =\frac{S_k(q+\tfrac12,\,N-2k+1,\,1)}{S_k(\tfrac12,\,N-2k+1,\,1)}
 =\prod_{j=1}^{k}
 \frac{\Gamma(q+j-\tfrac12)\,\Gamma(N-k+j-\tfrac12)}
      {\Gamma(j-\tfrac12)\,\Gamma(q+N-k+j-\tfrac12)}.
 \label{eq:app-selberg-ratio}
\end{equation}
This is \cref{eq:R-general} with $m=k$. Both Selberg integrals are finite for the positive moments considered here. The endpoint $k=0$ is deterministic and follows from the empty-product convention.

For $k>N/2$, the parametrization \cref{eq:app-jacobi} no longer applies directly: the block has $2k-N$ of its positive singular values pinned at $1$ and only $N-k$ fluctuating positive singular values, each representing a twofold-degenerate singular value of the antisymmetric matrix. Pfaffian complementation supplies the extension. Indeed, for a pure state, $\Gam^2=-\id$ gives $\Gam^{-1}=-\Gam$ and $\det\Gam=1$; Jacobi's complementary-minor identity then yields
$\det\Gam_A=\det\Gam_{A^c}$, or equivalently
$\Pf(\Gam_A)^2=\Pf(\Gam_{A^c})^2$. Hence $Y_A=Y_{A^c}$ and
$\E[Y_A^q]=R_{N,N-k}(q)$, which is \cref{eq:R-general} with
$m=\min(k,N-k)$. The second endpoint, $k=N$, is again deterministic. Rearranging the Gamma functions for positive integer $q$ gives the manifestly complement-symmetric form \cref{eq:R-integer}; summing the $\binom{2N}{2k}$ equivalent strings in each even-weight sector gives \cref{eq:maj-all-q} for $q>0$.

Finally, for every positive integer $q$, the ratio of consecutive terms in the sector sum, with $R_{N,k}(q)$ in the form \cref{eq:R-integer}, is a rational function of $k$, so the sum can be written as the terminating hypergeometric series
\begin{equation}
 \E\zeta_q^{\mathrm{Maj}}
 =2^{-N}\;{}_qF_{q-1}\!\left(
 \begin{matrix}-N,\ \tfrac32,\ \tfrac52,\ \ldots,\ q-\tfrac12\\[1mm]
 -N-\tfrac12,\ -N-\tfrac32,\ \ldots,\ \tfrac32-N-q\end{matrix}
 \;\middle|\;(-1)^q\right),
 \label{eq:maj-hypergeom}
\end{equation}
where the arithmetic parameter progressions are understood as empty when $q=1$. The upper parameter $-N$ truncates the series at degree $N$; we use this representation only at positive integer $q$ . The positive sector sum is generally preferable for numerical evaluation.

\section{The averaged spectral density and its cumulants}
\label{app:spectrum}

The beta-product law \cref{eq:beta-product} follows by Mellin inversion of \cref{eq:app-selberg-ratio}. Each factor in \cref{eq:R-general} is the Mellin transform of an independent $\operatorname{Beta}(j+\tfrac12,\,N-m)$ variable:
\begin{equation}
 \E[B_j^q]
 =\frac{\Gamma(j+\tfrac12+q)\,\Gamma(N-m+j+\tfrac12)}
        {\Gamma(j+\tfrac12)\,\Gamma(N-m+j+\tfrac12+q)}
 =\frac{\rising{j+\tfrac12}{q}}{\rising{N-m+j+\tfrac12}{q}}.
 \label{eq:app-beta-mellin}
\end{equation}
The product $\prod_{j=0}^{m-1}B_j$ therefore has moment $R_{N,k}(q)$ for every $q>0$. Since the variables are supported on the compact interval $[0,1]$, their nonnegative integer moments already determine the law uniquely, proving \cref{eq:beta-product}. Summing the sector laws with their multiplicities and restoring the delta peak from the vanishing odd-weight strings gives \cref{eq:maj-spectrum}.

Products of independent beta variables have Meijer $G$-function densities~\cite{SpringerThompson1970}. In the present parametrization, for $m\geq1$ and $0<y<1$,
\begin{equation}
 f_{N,m}(y)=
 \Biggl[\prod_{j=0}^{m-1}
 \frac{\Gamma(N-m+j+\tfrac12)}{\Gamma(j+\tfrac12)}\Biggr]\,
 G^{m,0}_{m,m}\!\left(y\,\middle|\,
 \begin{matrix}\{N-m+j-\tfrac12\}_{j=0}^{m-1}\\[1mm]
 \{j-\tfrac12\}_{j=0}^{m-1}\end{matrix}\right).
 \label{eq:meijer-density}
\end{equation}
Its Mellin transform is the product over $j$ of \cref{eq:app-beta-mellin}; the density integrates to one.

The logarithmic spectrum is controlled directly by the cumulants of
$\ln Y=\sum_j\ln B_j$. Conditional on Majorana weight $2k$, for every integer $p\geq1$,
\begin{equation}
 \kappa_p\bigl(\ln Y\mid k\bigr)
 =\sum_{j=0}^{m-1}
 \Bigl[\psi_{p-1}\bigl(j+\tfrac12\bigr)
 -\psi_{p-1}\bigl(N-m+j+\tfrac12\bigr)\Bigr],
 \label{eq:log-cumulants}
\end{equation}
where $\psi_p$ denotes the polygamma function of order $p$. Uniformly in every binomial central window $|k-N/2|\leq C\sqrt N$, with fixed $C$, these sums give
$\kappa_1=-N\ln2-\tfrac12\ln N+O(1)$ and
$\kappa_2=\ln N+O(1)$, while $\kappa_p=O(1)$ for every fixed $p\geq3$. Consequently, for
\begin{equation}
 Z_{N,k}:=\frac{\ln Y-\kappa_1}{\sqrt{\kappa_2}},
\end{equation}
every fixed standardized cumulant of order $p\geq3$ is
$O((\ln N)^{-p/2})$.
The method of moments gives $Z_{N,k}\Rightarrow\mathcal N(0,1)$: its moments converge to the Gaussian moments, which determine the limiting law. This proves that $Y$ is asymptotically lognormal in the central window, with mean and variance given by \cref{eq:central-lognormal}. The argument does not constrain the far tails, which control the high moments.

\bibliographystyle{jsty3-author}
\bibliography{fermionic_pauli_refs}
\end{document}